# New constraints on the galactic ionising efficiency and escape fraction at $2.5 < z < 6$ based on quasar absorption spectra

Christopher Cain,[1] Anson D'Aloisio,[2] and Julian B. Muñoz[3]

[1]School of Earth and Space exploration, Arizona State University, Tempe, AZ 85281, USA
[2]Department of Physics and Astronomy, University of California, Riverside, CA 92521, USA
[3]Department of Astronomy, The University of Texas at Austin, 2515 Speedway, Stop C1400, Austin, TX 78712, USA
**Author for correspondence:** Christopher Cain, Email: clcain3@asu.edu.

**Abstract**

Measurements of the ionisation state of the intergalactic medium (IGM) can probe the sources of the extragalactic ionising background. We provide new measurements of the ionising emissivity of galaxies using measurements of the ionising background and ionising photon mean free path from high-redshift quasar spectra at $2.5 < z < 6$. Unlike most prior works, we account for radiative-transfer effects and possible neutral islands from the tail of reionisation at $z > 5$. We combine our results with measurements of the UV luminosity function to constrain the average escaping ionising efficiency of galaxies, $\langle f_{\rm esc}\xi_{\rm ion}\rangle_{L_{\rm UV}}$. Assuming galaxies with $M_{\rm UV} < -11$ emit ionising photons, we find $\log(\langle f_{\rm esc}\xi_{\rm ion}\rangle_{L_{\rm UV}}/{\rm erg^{-1}Hz}) = 24.47^{+0.09}_{-0.17}$ and $24.75^{+0.15}_{-0.28}$ at $z = 5$ and 6, and $1\sigma$ upper limits of 24.48 and 24.31 at $z = 2.5$ and 4, respectively. We also estimate the population-averaged $f_{\rm esc}$ using measurements of intrinsic ionising efficiency from JWST. We find $\langle f_{\rm esc}\rangle = 0.126^{+0.034}_{-0.041}$ and $0.224^{+0.098}_{-0.108}$ at $z = 5$ and 6, and $1\sigma$ upper limits of $f_{\rm esc} < 0.138$ and 0.096 at $z = 2.5$ and 4, respectively, for $M_{\rm UV} < -11$. Our findings are consistent with prior measurements of $f_{\rm esc} \lesssim 10\%$ at $z \leq 4$, but indicate a factor of several increase between $z = 4$ and 6. The steepness of this evolution is sensitive to the highly uncertain mean free path and ionising background intensity at $z > 5$. Lastly, we find $1.10^{+0.21}_{-0.39}$ photons per H atom are emitted into the IGM between $z = 6$ and = 5.3. This is $\approx 4\times$ more than needed to complete the last 20% of reionisation absent recombinations, suggesting that reionisation's end was likely absorption-dominated.

## 1. Introduction

Despite recent advances in our understanding of the abundances and properties of high-redshift galaxies (e.g. Eisenstein et al. 2023; Adams et al. 2024; Finkelstein et al. 2024; Donnan et al. 2024; Harikane et al. 2024), their ionising properties remain highly uncertain. Recent efforts have focused on directly measuring $\xi_{\rm ion}$ using JWST (Simmonds et al. 2023; Atek et al. 2024; Pahl et al. 2024; Simmonds et al. 2024), and studying $f_{\rm esc}$ in low-redshift analogs of reionisation-era galaxies. One goal of the latter is to discover correlations with other observables that can be used to infer $f_{\rm esc}$ at higher redshifts (Pahl et al. 2021; Chisholm et al. 2022; Flury et al. 2022; Jaskot et al. 2024a, 2024b). However, a complementary approach is to measure the collective ionising output of galaxies independently by leveraging constraints on the ionisation state of the IGM, and then infer the required ionising properties of galaxies (e.g. Becker and Bolton 2013; Becker et al. 2021; Gaikwad et al. 2023; Bosman and Davies 2024).

The net ionising photon emissivity of the galaxy population is given by

$$\dot{N}_{\rm ion} = \rho_{\rm UV}\langle f_{\rm esc}\xi_{\rm ion}\rangle_{L_{\rm UV}} \quad (1)$$

where $\rho_{\rm UV}$ is the integrated UV luminosity density. The quantity $\langle f_{\rm esc}\xi_{\rm ion}\rangle_{L_{\rm UV}}$ is the UV-luminosity ($L_{\rm UV}$)-weighted product of two quantities that together quantify the ionising properties of galaxies. The first, $\xi_{\rm ion}$, is the "intrinsic ionising efficiency", or the rate of HI-ionising photon production per unit UV luminosity. The second, $f_{\rm esc}$, is the HI ionising escape fraction, which is the fraction of ionising photons produced within a galaxy that escape into the IGM. The galaxy UV luminosity function (UVLF) has been measured using JWST up to $z = 14$ (e.g. Pérez-González et al. 2023; Adams et al. 2023; Donnan et al. 2024; Harikane et al. 2024), facilitating direct measurements of $\rho_{\rm UV}$ up to that redshift. As such, an independent measurement of $\dot{N}_{\rm ion}$ using constraints on the IGM ionisation state would enable a measurement of $\langle f_{\rm esc}\xi_{\rm ion}\rangle_{L_{\rm UV}}$. One such approach is to combine information about the photoionisation rate in the ionised IGM inferred from the Ly$\alpha$ forest of high-redshift quasars (Wyithe and Bolton 2011; Calverley et al. 2011; Becker and Bolton 2013; D'Aloisio et al. 2018; Bosman et al. 2022) with measurements of the mean free path (MFP) to ionising photons (Prochaska, Worseck, and O'Meara 2009; Worseck et al. 2014; Becker et al. 2021; Zhu et al. 2023). The former contains information about the number of ionising photons in the IGM, and the latter about how quickly those photons are being absorbed. They can be combined to indirectly measure $\dot{N}_{\rm ion}$ at $z \leq 6$, where measurements of both quantities are available.

The exercise described above has been carried out by a number of previous works in the context of both HI and He II reionisation (e.g. Bolton and Haehnelt 2007; Kuhlen and Faucher-Giguère 2012; Becker and Bolton 2013; Khaire 2017)[a]. Most recently, Bosman and Davies (2024) (hereafter B24) used measurements of $\rho_{\rm UV}$ from Bouwens et al. (2021) to measure $\langle f_{\rm esc}\xi_{\rm ion}\rangle_{L_{\rm UV}}$ given an updated measurement of $\dot{N}_{\rm ion}$ at $z = 4-6$ by Gaikwad et al. (2023). Assuming galaxies produce ionising photons down to a limiting magnitude of $M_{\rm UV} = -11$, they found $\log\langle f_{\rm esc}\xi_{\rm ion}\rangle_{L_{\rm UV}} = 24.28^{+0.21}_{-0.20}$ erg$^{-1}$Hz at $z = 5$ and $\log\langle f_{\rm esc}\xi_{\rm ion}\rangle_{L_{\rm UV}} = 24.66^{+0.18}_{-0.47}$ erg$^{-1}$Hz at $z = 6$. They also report an upper limit at $z = 4$ of $\log\langle f_{\rm esc}\xi_{\rm ion}\rangle_{L_{\rm UV}} < 24.11$ erg$^{-1}$Hz. Notably, these measurements are up to a factor of $\sim 5$ lower than estimated in Muñoz et al. (2024) (for $M_{\rm UV} < -11$, Figure 1 of B24) using direct measurements of $\xi_{\rm ion}$ from JWST (Simmonds et al. 2023) and inferences on $f_{\rm esc}$ derived from low-redshift measurements (Chisholm et al. 2022) , suggesting a downward revision on the photon budget in the late stages of reionisation.

The measurements of $\dot{N}_{\rm ion}$ by B24, and most other works at these redshifts, neglect the effects of radiative transfer (RT) and reionisation. They rely on the so-called "local source approximation" (LSA), which assumes that the absorption rate of ionising photons in the IGM is equal to the emission rate. This is a good approximation during the bulk of reionisation, when the MFP is very short owing to much of the IGM being significantly neutral. However, at $5 < z < 6$, when reionisation is likely in its ending stages (Kulkarni et al. 2019; Keating, Kulkarni, et al. 2020; Nasir and D'Aloisio 2020), both the ionising background and the MFP evolve very rapidly. At this point, photons may experience a significant delay between emission and absorption. In this regime, an explicit treatment of RT effects may be required to accurately recover $\dot{N}_{\rm ion}$.

Another effect is that of possible "islands" of neutral hydrogen, which may still be present in the IGM down to redshifts as low as $z = 5$. Typically, measurements of $\dot{N}_{\rm ion}$ assume that the IGM is fully ionised. However, if neutral islands are present at $z > 5$, these likely affect measurements of the MFP (Roth et al. 2024; Satyavolu et al. 2024; Chen, Fan, and Avestruz 2024), and possibly the ionising background. This is of particular concern at $z = 6$, when reionisation was likely still ongoing and the global neutral fraction may be as high as 20% (Zhu, Becker, et al. 2024; Spina et al. 2024). As such, accounting for the presence of neutral islands at $z = 6$ may be necessary.

Measurements of $\dot{N}_{\rm ion}$ at these redshifts provide valuable insight into the evolution of galaxy properties and the tail end of the reionisation process. Recently, JWST has begun to measure $\xi_{\rm ion}$ for a large sample of galaxies up to $z \approx 9$ (e.g. Simmonds et al. 2023; Zhu, Alberts, et al. 2024). For the first time, these measurements statistically characterise the dependence of $\xi_{\rm ion}$ on both redshift and $M_{\rm UV}$ at redshifts extending into the reionisation epoch. Combining these new results with forest-based measurements of $\dot{N}_{\rm ion}$ will allow us to measure the average escape fraction of the galaxy population up to $z = 6$ more robustly than has been previously possible. Measuring the ionising output of galaxies at $z > 5$ will also begin to constrain the number of ionising photons required to complete reionisation - the ionising photon budget. The photon budget is sensitive to the ionising opacity of the ionised IGM, a major source of uncertainty in reionisation studies (So et al. 2014; Becker et al. 2021; Cain et al. 2021; Davies et al. 2021).

a. See also (e.g. Mason et al. 2019) for an example of constraining $\dot{N}_{\rm ion}$ at redshifts beyond those probed by the Ly$\alpha$ forest.

In this work, we present updated measurements of $\dot{N}_{\rm ion}$ at $2.5 \leq z \leq 6$, taking into account RT effects, and present new formalism to account for the presence of neutral islands in measurements of $\dot{N}_{\rm ion}$. We interpret our findings in the context of high-redshift galaxy observations, and infer estimates of the escaping ionising efficiency and the population-averaged escape fraction. This work is organised as follows. In Section 2, we present the formalism used to measure $\dot{N}_{\rm ion}$. We describe the observations used in our analysis and our modelling uncertainties in Section 3, present our main results in Section 4, and conclude in Section 5. Throughout, we assume the following cosmological parameters: $\Omega_m = 0.305$, $\Omega_\Lambda = 1 - \Omega_m$, $\Omega_b = 0.048$, $h = 0.68$, $n_s = 0.9667$ and $\sigma_8 = 0.82$, consistent with Planck Collaboration et al. 2020 results. All distances are co-moving unless otherwise specified.

## 2. Formalism

### 2.1 Radiative transfer equation

In the LSA, the IGM photo-ionisation rate $\Gamma_{\rm HI}$ can be expressed as

$$\Gamma_{\rm HI} = (1+z)^3 \int d\nu \dot{N}^{\nu}_{\rm ion}\lambda_\nu \sigma^{\nu}_{\rm HI} \tag{2}$$

where $\dot{N}^{\nu}_{\rm ion}$ is the co-moving ionising emissivity output by sources per unit frequency $\nu$, $\lambda_\nu$ is the MFP to ionising photons, and $\sigma^{\nu}_{\rm HI}$ is the HI-ionising cross-section. Equation (2) supposes that (1) photons are absorbed within a short time after being emitted and (2) that the IGM is highly ionised. A more general expression that does not rely on the first assumption is

$$\Gamma_{\rm HI} = (1+z)^3 \int d\nu \dot{N}^{\nu}_{\rm abs}\lambda_\nu \sigma^{\nu}_{\rm HI} \tag{3}$$

where we have replaced the emission rate $\dot{N}^{\nu}_{\rm ion}$ with the absorption rate, $\dot{N}^{\nu}_{\rm abs}$, no longer assuming these to be equal. This distinction is important whenever the timescale for an ionising photon to travel one mean free path, $\lambda_\nu/c$, is significant compared to the timescales over which IGM properties evolve. The absorption rate at time $t$ is given by

$$\dot{N}^{\nu}_{\rm abs}(t) = \int_0^t dt' \dot{N}^{\nu}_{\rm ion}(t',\nu')G(t,t',\nu,\nu') \tag{4}$$

where photons emitted at time $t' \leq t$ ($z' \geq z$) with frequency $\nu' = \frac{1+z'}{1+z}\nu$ red-shifting to observed frequency $\nu$ at time $t$, and $\dot{N}^{\nu}_{\rm ion}(t',\nu')$ is the emissivity per unit co-moving frequency ($\nu$) evaluated at $\nu'$. Here, $G$ is given by

$$G(t,t',\nu,\nu') = c\kappa_\nu(t)\exp\left[-\int_{t'}^{t} dt'' c\kappa_{\nu''}(t'')\right] \tag{5}$$

where $\kappa_\nu \equiv \lambda_\nu^{-1}$ is the absorption coefficient, $c$ is the speed of light, and $\nu'' = \frac{1+z''}{1+z}\nu$. $G$ is the Green's function for the cosmological RT equation (e.g. in Haardt and Madau 1996, 2012) with the usual proper emissivity and angle-averaged intensity replaced by the co-moving ionising photon emissivity and absorption rate, respectively. It quantifies how the ionising background at time $t$ is built up by photons emitted at earlier times, $t' < t$. In the limit that $1/(c\kappa_\nu) \to 0$ (the short MFP limit), $G(t, t', \nu, \nu')$ becomes the Dirac delta function $\delta_D(t - t', \nu - \nu')$, in which case we recover the LSA. We show in Appendix 1 that our formulation is equivalent to the usual way of expressing the solution to the cosmological RT equation. The advantage to our formulation is that Eqs. (3-5) show directly the relationships between $\lambda$, $\Gamma_{HI}$, and $\dot{N}_{ion}$.

### 2.2 Accounting for neutral islands

Equation (3) is only accurate if all photons are absorbed by the ionised IGM and contribute to its total ionisation rate. Photons absorbed by any neutral islands will not contribute to $\Gamma_{HI}$ in the ionised IGM, which is what the Ly$\alpha$ forest probes. As such, it is unclear whether the standard approach to measuring $\dot{N}_{ion}$ will give the right answer in an IGM containing neutral islands. A modified version of Equation (3) accounting for this is

$$\Gamma_{HI} = \int d\nu \dot{N}^{\nu}_{abs,ionised} \lambda_{\nu,ionised} \sigma^{\nu}_{HI} \tag{6}$$

where $\dot{N}^{\nu}_{abs,ionised}$ is the absorption rate in ionised gas, and $\lambda^{ionised}_{\nu}$ is the MFP in ionised gas. These quantities are defined such that the contribution of neutral islands to both is removed. We can estimate $\lambda_{\nu,ionised}$ by

$$\lambda^{-1}_{\nu,ionised} = \lambda^{-1}_{\nu} - \lambda^{-1}_{\nu,neutral} \tag{7}$$

where $\lambda^{-1}_{\nu,neutral}$ is the contribution to the absorption coefficient from neutral islands, i.e. not counting the opacity of the ionised gas. Similarly, we can estimate $\dot{N}^{\nu}_{abs,ionised}$ using

$$\dot{N}^{\nu}_{abs,ionised} = \dot{N}^{\nu}_{abs} - \dot{N}^{\nu}_{abs,neutral} \tag{8}$$

where the absorption rate by neutral islands only is given by

$$\int d\nu \dot{N}^{\nu}_{abs,neutral} = -\frac{dx^{m}_{HI}}{dt}(1 + \chi) n_H \tag{9}$$

Here, $x^{m}_{HI}$ is the mass-weighted HI fraction in the IGM, $n_H$ is the cosmic mean hydrogen density, and the factor of $1 + \chi \equiv 1.082$ accounts for single ionisation of He. Equation (9) simply counts the net rate at which the fully neutral IGM is being ionised, which is equivalent to the absorption rate by neutral islands only. Since the global reionisation history is unknown, in our analysis we must assume it (and the MFP to neutral islands, $\lambda_{\nu,neutral}$) from a simulation, which we discuss further in the next section. We explore the effect of accounting for neutral islands in this way in Appendix 2.

In sum, our RT formalism takes into account several effects that are missing in the LSA. First, it accounts for red-shifting of ionising photons past the Lyman Limit, which becomes important at $z \lesssim 4$ (Becker and Bolton 2013). Second, it accounts for the rapid buildup of the ionising background immediately following reionisation, which necessitates $\dot{N}^{\nu}_{ion} > \dot{N}^{\nu}_{abs}$. Lastly, including neutral islands accounts for the possibility that reionisation may be ongoing, such that not all absorptions occur in the highly ionised IGM.

## 3. Method

### 3.1 Measurements of $\Gamma_{HI}$ and $\lambda$

We use measurements of $\Gamma_{HI}$ and the Lyman Limit MFP, $\lambda^{mfp}_{912}$, to measure $\dot{N}_{ion}$ at $2.5 < z < 6$. We show these measurements in Figure 1 - see the legend and caption for references[b]. In practice, solving Equation (3-5) for $\dot{N}_{ion}$ requires a smooth functional form for both $\Gamma_{HI}$ and $\lambda^{mfp}_{912}$ (see Section 3.2). We fit the collection of measurements to smooth functions of redshift using an MCMC approach, accounting for the reported $1\sigma$ error bars on the measurements using a standard Gaussian likelihood[c]. We use a 5th-order polynomial in redshift to fit the $\log_{10}\Gamma_{HI}$ and a double power-law to fit the $\lambda^{mfp}_{912}$ measurements (see Appendix 3 for details). The maximum likelihood fit is shown by the red dashed curve in each panel of Figure 1. The thin black lines are random draws from the recovered posteriors on the best-fit parameters, which approximately capture the error in the best-fits to the measurements (see Section 3.3). The cyan dot-dashed curve in the bottom panel shows the best fit with the neutral island opacity subtracted (Equation (7), and see below).

A key source of uncertainty is the choice of $\Gamma_{HI}$ and $\lambda^{mfp}_{912}$ measurements used at $z \geq 5$. For both $\Gamma_{HI}$ and $\lambda^{mfp}_{912}$, the measurements in Figure 1 represent two different approaches to measuring these quantities. The standard way to measure $\Gamma_{HI}$ is to match the mean transmission of the Ly$\alpha$ forest in hydrodynamical simulations with that measured in the spectra of high-redshift quasars. Typically, these simulations assume that reionisation is over and that the UV background (UVB) is homogeneous. Significant sources of uncertainty in such measurements include the thermal history of the IGM, which is typically marginalised over (Becker and Bolton 2013), and numerical convergence in both spatial resolution and box size (Doughty et al. 2023). The measurements of Becker and Bolton (2013), Bosman et al. (2018)[d], and Becker et al. (2021) in the top panel were done in this way. An important caveat is that the Bosman et al. (2022) measurements were done assuming an early reionisation model with a relatively low IGM temperature, and thus may be biased high (see below).

Direct measurements of $\lambda^{mfp}_{912}$ use the Lyman Continuum (LyC) spectrum of high-redshift quasars. The observed MFP

b. Note that we exclude the MFP measurements of Becker et al. (2021) at $z = 5.1$ and 6 because they are measured using a similar data set and method as those from Zhu et al. (2023).

c. Asymmetric error bars are accounted for in our likelihood using the variable Gaussian approach described in Barlow (2004).

d. The measurements of Bosman et al. (2022) do not include error bars. We assigned error bars to these measurements to roughly match those of the G23 points. These are consistent with the expected factor of $\sim 2$ level uncertainty in the IGM thermal history (D'Aloisio et al. 2018).

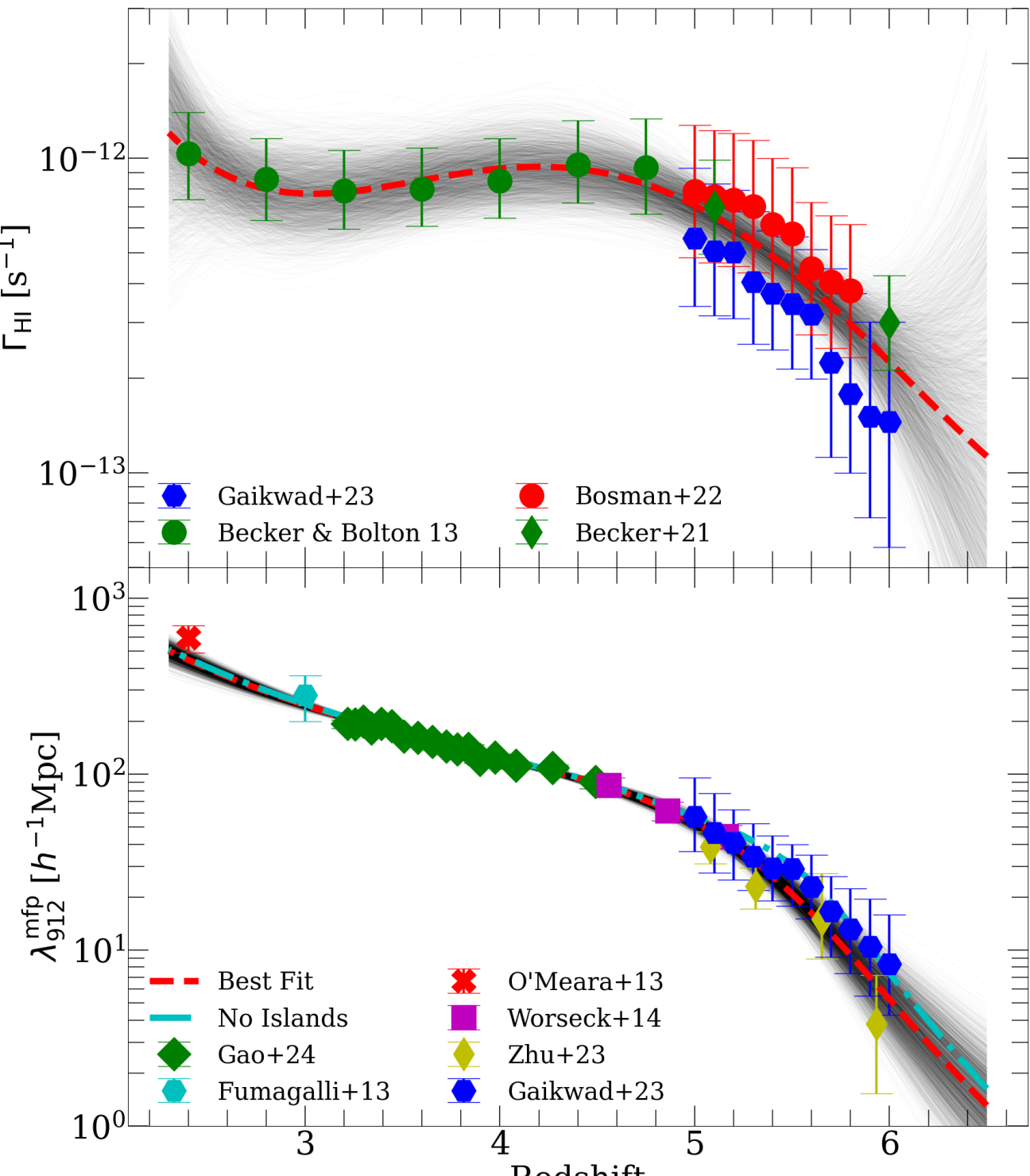

**Figure 1.** Collection of measurements of $\Gamma_{\rm HI}$ (top) and $\lambda_{912}^{\rm mfp}$ (bottom) used to measure $\dot{N}_{\rm ion}$ in this work. The red dashed curves shows the maximum-likelihood fits to each set of measurements. The thin black curves show random draws from the posteriors of the model parameters. The cyan dot-dashed curve in the bottom panel shows the "ionised phase" MFP estimated using Equation (7). Measurements of $\Gamma_{\rm HI}$ are from Becker and Bolton (2013), Bosman et al. (2022), and Gaikwad et al. (2023). Measurements of $\lambda_{912}^{\rm mfp}$ are from Fumagalli et al. (2013), O'Meara et al. (2013), Worseck et al. (2014), Zhu et al. (2023), Gaikwad et al. (2023), and Gao et al. (2024).

at a given redshift is defined as the distance travelled by a photon emitted at the redshift of the quasar that reaches a LyC opacity of unity when it redshifts to 912Å. This quantity is estimated by stacking and fitting LyC spectra of bright quasar spectra (Prochaska, Worseck, and O'Meara 2009). All the measurements in Figure 1 use this approach except those of Gaikwad et al. (2023, see below). One important caveat is that measurements using this method at $z > 5$ are challenging and must account for the effect of the proximity zone of the quasar on the LyC spectrum. This may introduce additional model-dependence and potential systematic uncertainty in the measurement (Becker et al. 2021).

Recently, Gaikwad et al. (2023, hereafter G23) proposed a new method of jointly measuring $\Gamma_{\rm HI}$ and $\lambda_{912}^{\rm mfp}$ based on the fluctuations in the Ly$\alpha$ forest on large spatial scales (see also Davies et al. 2024). Their approach is somewhat complementary to the standard approaches, since it uses different information available in the quasar spectra. It relies on semi-numerical simulations that assume a relationship between $\lambda_{912}^{\rm mfp}$ and large-scale fluctuations in both $\Gamma_{\rm HI}$ and the IGM density, which makes it (to some degree) model-dependent. Crucially, however, it takes into account spatial fluctuations in the UVB and the possible presence of neutral islands near the end of reionisation ($z > 5$). Their measurements are the blue points in both panels of Figure 1. Note that all the most recent $z > 5$ measurements of both $\Gamma_{\rm HI}$ and $\lambda_{912}^{\rm mfp}$ use quasar spectra from the XQR-30 data set (D'Odorico et al. 2023).

To gauge the differences between $\dot{N}_{\rm ion}$ implied by these two types of measurements at $z > 5$, we also fit two sub-sets of the data. The first includes only the G23 measurements (for both quantities) at $z \geq 5$ and the second excludes only the G23 results. Both sub-sets include the $\Gamma_{\rm HI}$ measurements of Becker and Bolton (2013) at $z < 5$ and all direct $\lambda_{912}^{\rm mfp}$ measurements at $z < 5$. These fits are shown in Appendix 3 and are used in the analysis below. The parameters for the maximum-likelihood fits to our full data set and these two reduced data sets are also in Appendix 3. We note that differences in $\Gamma_{\rm HI}$ and $\lambda_{912}^{\rm mfp}$ between the two sub-sets of measurements contribute at similar levels to differences in measured quantities at $z > 5$.

### 3.2 Estimation of $\dot{N}_{\rm ion}$

To measure $\dot{N}_{\rm ion}$ from $\Gamma_{\rm HI}$ and $\lambda_{912}^{\rm mfp}$ we need to the frequency ($\nu$) dependence of the spectrum of the emitted ionising radiation, and the frequency dependence of $\lambda_\nu$. We model the former as a power law in $\nu$, such that

$$\dot{N}_{\rm ion}^{\nu} \propto \frac{\nu^{-\alpha}}{h_p \nu} \propto \nu^{-(\alpha+1)} \tag{10}$$

where $h_p$ is Planck's constant, for photon energies between 1 and 4 Ryd. The cutoff at $E > 4$ Ryd is appropriate for galaxy spectra, and a reasonable approximation for quasars prior to the onset of He II reionisation. At $z \lesssim 4$, when He II reionisation is likely underway, emission at energies greater than 4 Ryd from quasars contributes to the ionising background in most of the IGM. We note, however, that this effect on our $\dot{N}_{\rm ion}$ measurements is < 15% over most of our parameter space, well below the total error budget (see next section)[e].

Assuming the HI column density distribution has the form of a power law with slope $\beta_N$, we can write (McQuinn, Oh, and Faucher-Giguère 2011),

$$\lambda_\nu \propto (\sigma_{\rm HI}^{\nu})^{-(\beta_N - 1)} \propto \nu^{2.75(\beta_N - 1)} \tag{11}$$

where we have approximated $\sigma_{\rm HI} \propto \nu^{-2.75}$ in the frequency range of interest. The limit that $\beta_N = 1$ (no frequency dependence) is that of the IGM opacity being dominated by highly opaque ($\tau >> 1$) gas. In the opposite limit ($\beta_N = 2$), it is dominated by the diffuse, highly ionised IGM. Smaller $\alpha$ corresponds to harder (more energetic) ionising spectra,

e. Assuming quasar spectra have the same power law spectral shape above and below 4 Ryd, and 100% of the ionising budget comes from quasars, the error incurred in Equation 2 is a factor of $(1 - 4^{-\alpha+2.75(\beta_N-2)})/(1 - 4^{-\alpha})$. For our fiducial choice of $\alpha = 2$, $\beta_N = 1.3$, this error is $\approx 6\%$. If $\alpha = 1$, this error increases to $\approx 30\%$ for $\beta_N \lesssim 1.5$. In reality, however, during He II reionisation a significant fraction of the > 4 Ryd photons will be absorbed by He II, decreasing this error in Equation 3.

which results in lower $\Gamma_{\rm HI}$ at fixed $\dot{N}_{\rm ion}$ because of the frequency dependence of $\sigma_{\rm HI}$. Thus, assuming a smaller $\alpha$ requires a larger $\dot{N}_{\rm ion}$ at fixed $\Gamma_{\rm HI}$. A smaller $\beta_{\rm N}$ decreases the frequency-averaged MFP at fixed $\lambda_{912}^{\rm mfp}$, which also increases $\dot{N}_{\rm ion}$ at fixed $\Gamma_{\rm HI}$. Following B24, we assume fiducial values of $\alpha = 2$ and $\beta_{\rm N} = 1.3$. We will vary these parameters in the ranges $1 \leq \alpha \leq 3$ and $1 \leq \beta_{\rm N} \leq 2$. Note that our choice of $\alpha = 2$ is motivated by models of metal-poor (Pop II) galaxy SEDs (Bressan et al. 2012; Choi, Conroy, and Byler 2017), but is also reasonable for quasars (Lusso et al. 2015), which probably dominate the ionising output of the source population at $z < 4$ (see, however, Madau et al. 2024 for higher $z$). Our choice of $\beta_{\rm N}$ is commonly assumed in the literature (e.g. Gaikwad et al. 2023) and is motivated by the best-fit to the HI column density distribution in Becker and Bolton (2013).

The integration in Equation (4) runs over all $t' < t$, so formally it should start from $z = \infty$ ($t = 0$). In practice, it is not possible to do this, since we have measurements of $\Gamma_{\rm HI}$ and $\lambda_{912}^{\rm mfp}$ only up to $z = 6$. To approximate Equation (4), we extrapolate our maximum-likelihood fits for $\Gamma_{\rm HI}$ and $\lambda_{912}^{\rm mfp}$ to $z = 6.5$, and assume that at that redshift the LSA is valid. This lets us estimate $\dot{N}_{\rm ion}(z = 6.5)$ using Equation (2). We then set the lower limit of the integration in Equation (4) to $t(z = 6.5)$ and adjust the value of $\dot{N}_{\rm ion}$ at each successively lower redshift until Equation (3) returns the measured $\Gamma_{\rm HI}$. In this way, we find the $\dot{N}_{\rm ion}(z)$ down to $z = 2.5$ that satisfies Equation (3-5) for our assumed $\Gamma_{\rm HI}$ and $\lambda_{912}^{\rm mfp}$. Our results at $2.5 \leq z \leq 6$ are only sensitive at the few-percent level (or less) to how $\Gamma_{\rm HI}$ and $\lambda_{912}^{\rm mfp}$ are extrapolated[f] to $z = 6.5$.

Since reionisation may be ongoing at $z > 5$, we also use Equation (6-9) to measure $\dot{N}_{\rm ion}$ assuming a late reionisation history ending at $z \approx 5$. To do this, we need (1) a reionisation history, to evaluate Equation (9), (2) a neutral island MFP to evaluate Equation (7) and (3) a functional form for the spectrum of ionising radiation absorbed by neutral islands (to extract $\dot{N}_{\rm abs,neutral}^{\nu}$ from the integral in Equation (9)). To satisfy (1) and (2), we use the LATE START/LATE END model from Cain, Lopez, et al. (2024), which is a ray-tracing RT simulation of the EoR run with the FlexRT code of Cain and D'Aloisio (2024). The $\dot{N}_{\rm ion}$ in the simulation is calibrated to produce agreement with the mean Ly$\alpha$ forest transmission measurements of Bosman et al. (2022) at $5 \leq z \leq 6$. It has a volume-averaged neutral fraction of 30% at $z = 6$, making it a fairly extreme case of late reionisation (for discussion, see Zhu, Becker, et al. 2024). As such, our estimate of the effect of neutral islands on measurements of $\dot{N}_{\rm ion}$ is likely an over-estimate, to be viewed as a rough upper limit. However, by comparing models with and without neutral islands, we will see that they have only a mild effect on our results. We evaluate the MFP to neutral islands in the simulation by setting the opacity in ionised gas to 0 and then estimating $\lambda$ using the definition described in Appendix C of Chardin et al. 2015 (Equation 5 in Cain, Lopez, et al. (2024)). For (3), we assume (for simplicity) that the the spectrum of $\dot{N}_{\rm abs,neutral}^{\nu}$ is the same[g] as that of $\dot{N}_{\rm abs}^{\nu}$, allowing us to straightforwardly evaluate Equation (8).

### 3.3 *Errors on* $\dot{N}_{\rm ion}$

Errors on $\dot{N}_{\rm ion}$ arise from uncertainty in $\Gamma_{\rm HI}$, $\lambda_{912}^{\rm mfp}$, $\alpha$, and $\beta_{\rm N}$. To estimate uncertainties on $\dot{N}_{\rm ion}$, we draw 1000 random sets of parameters from the posteriors on the MCMC fits to $\Gamma_{\rm HI}$ and $\lambda_{912}^{\rm mfp}$ and calculate $\dot{N}_{\rm ion}$ for each combination. For each draw, we also randomly draw values of $\alpha$ and $\beta_{\rm N}$ from the uniform distributions $\alpha \in [1, 3]$ and $\beta_{\rm N} \in [1, 2]$. At each redshift, we treat the 13 – 87% (2.5 – 97.5%) range of the resulting $\dot{N}_{\rm ion}$ values as the $1\sigma$ ($2\sigma$) spread around our maximum likelihood fiducial measurement. Note that our $1\sigma$ and $2\sigma$ ranges are not, in general, symmetric around our maximum likelihood result[h]. We do this for our fiducial fit using all data points, and our reduced data sets including and excluding the G23 measurements.

To quantify how important each parameter is to determining the errors in $\dot{N}_{\rm ion}$, we re-run this analysis allowing only one parameter at a time to vary, holding all the others fixed to their maximum likelihood fits or fiducial values. In Table 1, we report the two-sided $1\sigma$ ($2\sigma$ in parentheses) logarithmic errors on $\dot{N}_{\rm ion}$ arising from each parameter individually at $z = 2.5$, 4, 5, and 6. The bottom row reports the total errors on $\dot{N}_{\rm ion}$. At $z = 2.5$, uncertainties in $\Gamma_{\rm HI}$ dominate the error budget. At $z = 4$ and 5, $\Gamma_{\rm HI}$ and $\beta_{\rm N}$ dominate, and contribute at roughly equal levels. At $z = 6$, uncertainties in $\lambda_{912}^{\rm mfp}$ become important at a level similar to $\Gamma_{\rm HI}$ and $\beta_{\rm N}$. Uncertainties from $\alpha$ are sub-dominate at all redshifts. The full redshift dependence of these one-parameter errors is shown in Appendix 4.

We mention here a couple of caveats that likely render our uncertainties on $\dot{N}_{\rm ion}$ too small. The first is that they rely on parametric fits to measurements of $\Gamma_{\rm HI}$ and $\lambda_{912}^{\rm mfp}$. These do not take into account the full covariance between uncertainties on different measurements, and the scatter in the posteriors is likely too small on account of the relatively small number of parameters (4–5) used in the fits. Another caveat is that the two different types of measurements discussed in Section 3.1 are systematically offset from each other at $z > 5$, as can be clearly seen in Figure 1. Using both in the MCMC fits will thus produce overly tight posteriors at $z > 5$. In what follows, we will show results using each set of measurements individually,

f. This is because the MFP is short enough at $z \geq 6$ that the LSA is still approximately valid.

g. This is not true in general, as the ionising photons reaching neutral islands may have experienced significant hardening by the IGM, and are thus on average more energetic than those absorbed in the ionised IGM (Wilson et al. 2024). However, since the simulations upon which our neutral island correction is based are monochromatic, this information is unfortunately not available. We also do not expect this approximation to meaningfully affect our measurements.

h. The largest source of this asymmetry is $\beta_{\rm N}$. The fiducial value of 1.3 gives $\dot{N}_{\rm ion}$ on the high end of the $[1, 2]$ range we marginalise over in the uncertainty calculation. Some additional asymmetry arises from the maximum likelihood fit to $\Gamma_{\rm HI}$ measurements not being exactly at the median of the posterior.

**Table 1.** Estimate of the error budget for our fiducial $\dot{N}_{\rm ion}$ measurement at several redshifts. The bottom row reports the total logarithmic errors on the measurements, and the rows above give an estimate of the contribution from each uncertain quantity in the analysis. We report $\pm 1\sigma$ errors and $2\sigma$ errors in parentheses.

| $z$ | 2.5 | 4 | 5 | 6 |
|---|---|---|---|---|
| $\Gamma_{\rm HI}$ | +0.101(0.23) | +0.084(0.16) | +0.055(0.11) | +0.107(0.22) |
| | -0.232(0.44) | -0.069(0.14) | -0.059(0.11) | -0.111(0.22) |
| $\lambda_{912}^{\rm mfp}$ | +0.012(0.02) | +0.007(0.01) | +0.021(0.04) | +0.125(0.25) |
| | -0.007(0.02) | -0.004(0.01) | -0.017(0.03) | -0.128(0.22) |
| $\alpha$ | +0.045(0.07) | +0.045(0.06) | +0.052(0.07) | +0.053(0.08) |
| | -0.031(0.04) | -0.030(0.04) | -0.039(0.05) | -0.043(0.06) |
| $\beta_{\rm N}$ | +0.021(0.04) | + 0.039(0.07) | + 0.036(0.06) | + 0.044(0.07) |
| | -0.056(0.07) | -0.129(0.17) | -0.120(0.16) | -0.168(0.22) |
| All | +0.100(0.22) | +0.080(0.17) | +0.054(0.15) | +0.117(0.32) |
| | -0.265(0.52) | -0.159(0.26) | -0.151(0.24) | -0.272(0.46) |

allowing us to quantify better the real uncertainties in $\dot{N}_{\rm ion}$ at these redshifts. We discuss these caveats in more detail in Appendix 4.

## 4. Results

### 4.1 Measurements of $\dot{N}_{\rm ion}$

We show our fiducial measurements of $\dot{N}_{\rm ion}$ at $2.5 < z < 6$ in the top panel of Figure 2. The solid black curve shows our max-likelihood measurement, and the shaded region indicates the $1\sigma$ uncertainties. Since our measurement and error bars at each redshift are a result of a cumulative integral over higher redshifts, we show them as a continuous curve rather than as discrete points at particular redshifts. We show, for comparison, the $\dot{N}_{\rm ion}$ measurements from G23 at $z = 5$ and 6 (blue points) and those by Becker and Bolton (2013) at $2 < z < 5$ (green points). Notably, our measurement is $\approx 0.8\sigma$ higher than that of G23 at $z = 5$ and $\approx 2\sigma$ higher at $z = 6$. At $z < 5$, our measurement is close to the central values of the Becker and Bolton (2013) measurements.

In our fiducial measurement, $\dot{N}_{\rm ion}$ falls by a factor of $\approx 2$ between $z = 6$ and 5. This is because the measured $\lambda_{912}^{\rm mfp}$ grows with time more quickly over that redshift range than $\Gamma_{\rm HI}$ (since $\dot{N}_{\rm ion} \sim \Gamma_{\rm HI}/\lambda_{912}^{\rm mfp}$). We find roughly constant $\dot{N}_{\rm ion}$ between $z = 5$ and 4, with a small dip around $z = 3-3.5$ and a sharp increase towards $z = 2.5$. This last feature is driven by the slight upturn in the Becker and Bolton (2013) $\Gamma_{\rm HI}$ measurements at $z < 3.5$[i]. Our $\pm 1\sigma$ range is a factor of $\approx 2$ at most redshifts, but increases considerably at $z > 5.5$ due to rising uncertainty in $\lambda_{912}^{\rm mfp}$ (see Table 1).

The red-dashed curve in Figure 2 shows the effect of using the LSA to compute $\dot{N}_{\rm ion}$ instead of the full formalism accounting for RT. At $z \approx 6$, this agrees well with the full measurement, thanks to the short MFP. However, at $z < 6$, they start to diverge, and by $z = 5$ the LSA gives a result $\approx 20\%$ below the full calculation. This effect explains over half the difference between our measurement and that of G23, and the rest arises from differences in the assumed $\Gamma_{\rm HI}$ and $\lambda_{912}^{\rm mfp}$. At $z < 5$, the difference between the LSA and the full measurement increases, reaching a factor of $\approx 3$ by $z = 2.5$. Previous works (including Becker and Bolton 2013) have noted that at $z < 4$, the LSA is expected to fail because it ignores the red-shifting of ionising photons past the Lyman Limit before they can ionize a neutral atom. The black dotted curve isolates this effect by explicitly neglecting the red-shifting of photons while still accounting for the distance they travel through the

i. Physically, this may be driven by the growth of the ionising output of quasars and the reionisation of He around these redshifts.

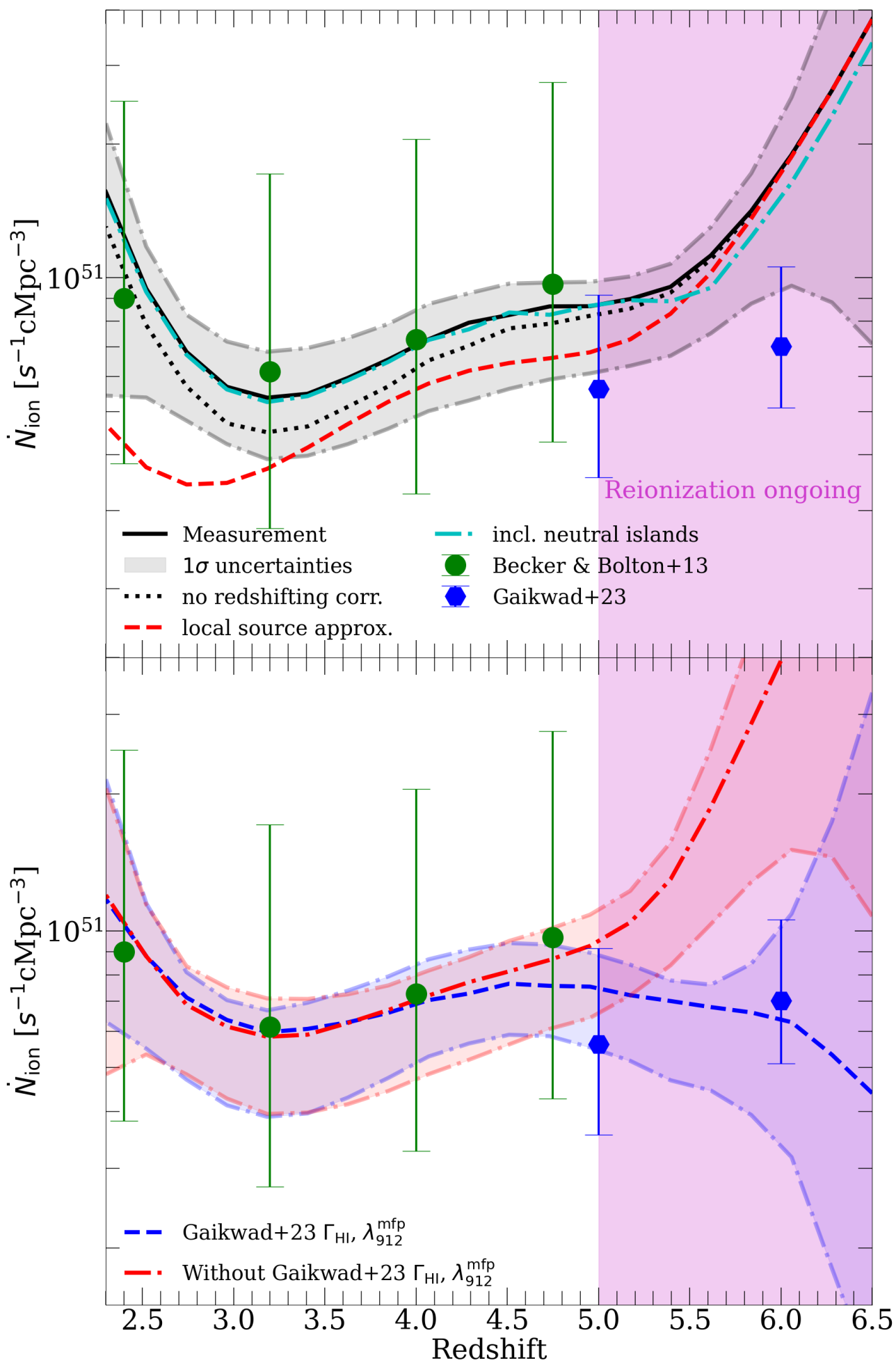

**Figure 2.** Measurements of $\dot{N}_{\rm ion}$ compared to previous measurements. **Top:** our fiducial measurement (black solid curve) at $2.5 < z < 6$. The dark (light) shaded region indicates the approximate $1\sigma$ ($2\sigma$) uncertainties. The red dashed curve shows estimates of $\dot{N}_{\rm ion}$ using the LSA, and the black dotted curve includes neglects red-shifting of ionising photons past the Lyman Limit, but still accounts for the finite travel time of photons. The cyan dot-dashed curve shows our revised measurement accounting for the presence of neutral islands at $z > 5$. **Bottom:** the same measurement (including only $1\sigma$ uncertainties) using our reduced data sets with only the G23 points at $z \geq 5$ (blue dashed curve) and the excluding only the G23 points (red dot-dashed curve). The differences between these are small at $z < 5$, but become significant at $z > 5$. In the former case, $\dot{N}_{\rm ion}$ remains nearly flat up to $z = 6$, while in the latter, $\dot{N}_{\rm ion}$ grows by a factor of $\approx 4$ between $z = 5$ and 6. In our fiducial measurement, this increase is a factor of $\approx 2$.

IGM before being absorbed[j]. This result differs from the full measurement by at most $\approx 25\%$, and explains less than half of the difference between the black solid and red dashed curves most of the time. This indicates that the steady buildup of ionising photons in the IGM due to the lengthening MFP contributes more to the failure of the LSA than does redshifting, even at low redshifts.

The cyan dot-dashed curve includes the effect of neutral islands at $z > 5$, when reionisation is ongoing in our assumed model (vertical shaded region). We see that taking islands into account changes the result by at most 15%. We can understand why by returning to Equation (6). In that equation, $\dot{N}^{\nu}_{\rm abs,ionised}$ is smaller than the total absorption rate, $\dot{N}^{\nu}_{\rm ion}$, since some ionising photons are consumed by the neutral islands (Equation (9)). However, the MFP in ionised gas is also larger than that in the IGM at large (Equation (7)). It turns out that the product $\dot{N}^{\nu}_{\rm abs,ionised}\lambda_{\nu,{\rm ionised}}$ is similar to $\dot{N}^{\nu}_{\rm abs}\lambda_{\nu}$, meaning we recover the nearly the same total $\dot{N}_{\rm ion}$ that we would if we had ignored islands. We investigate this result more carefully in Appendix 2, and find it to be true even when $\dot{N}_{\rm abs,neutral} > \dot{N}_{\rm abs,ionised}$ (that is, when islands dominate the global absorption rate).

The bottom panel of Figure 2 shows how our results change when only the G23 measurements (Figure 1) are used at $z \geq 5$ (blue dashed) and when they are excluded (red dot-dashed). While the results are similar at $z < 5$, they diverge considerably at $z > 5$. The maximum likelihood result using only the G23 measurements declines slightly with redshift at $z > 5$. However, without G23, we find a factor of $\approx 4$ increase in $\dot{N}_{\rm ion}$ from $z = 5$ to 6 in our maximum likelihood result, and scenarios with flat or decreasing $\dot{N}_{\rm ion}$ are clearly disfavoured. The spread between these two sets of results better captures the uncertainties in $\dot{N}_{\rm ion}$ at $z > 5$, which are artificially tightened when combining all measurements (see Section 3.3).

The differences between the blue and red curves arise from the fact that the $\Gamma_{\rm HI}$ measured by G23 decreases more quickly, and $\lambda^{\rm mfp}_{912}$ less quickly, with redshift than in the other subset of measurements. Notably, their measurement of $\lambda^{\rm mfp}_{912}$ at $z = 6$ is a factor of 2 higher than the direct measurements from Zhu et al. (2023). Their $z = 6$ $\Gamma_{\rm HI}$ measurement is also a factor of $\approx 2$ below that of Becker et al. (2021) at $z = 6$. In the G23-only subset, $\Gamma_{\rm HI}$ and $\lambda^{\rm mfp}_{912}$ decrease with redshift at the roughly the same rate, keeping $\dot{N}_{\rm ion}$ approximately constant. In the other subset, $\Gamma_{\rm HI}$ decreases much more slowly than $\lambda^{\rm mfp}_{912}$, causing $\dot{N}_{\rm ion}$ to grow rapidly. The difference shrinks to $\approx 20\%$ at $z = 5$ and disappears by $z = 4$.

### *4.2 ionising photon budget*

Several recent works have suggested, based on Ly$\alpha$ forest observations, that reionisation ended at $z = 5 - 5.5$. If this is the case, we can use our measurements to estimate the ionising photon output per H atom, $N_{\gamma/H}$, at the tail end of reionisation. Assuming reionisation ended at $z = 5.3$, as suggested by Bosman et al. (2022), we define the $5.3 < z < 6$ photon budget to be the number of ionising photons produced per H atom between $z = 5.3$ and 6, $N^{z=5.3-6}_{\gamma/H}$. We measure $N^{z=5.3-6}_{\gamma/H} = 1.10^{+0.21}_{-0.39}$ for our fiducial measurement at $1\sigma$ confidence, and $0.61^{+0.13}_{-0.22}$ ($1.95^{+0.93}_{-0.98}$) using our sub-sets of $\Gamma_{\rm HI}$ and $\lambda^{\rm mfp}_{912}$ with (without) the G23 results, respectively.

If the universe were 20% neutral at $z = 6$, as suggested by recent models, the minimum budget required to complete reionisation by $z = 5.3$ in the absence of recombinations is $N^{z=5.3-6}_{\gamma/H} = 0.216$ (which accounts for single ionisation of Helium). We can consider the tail-end of reionisation to be "absorption-dominated" if the actual budget is at least twice this value (see Davies et al. 2021), which is true even when using only G23 measurements at $z > 5$. In our most extreme scenario, the tail of reionisation is absorption-dominated by a factor of $\approx 9$. This suggests that absorption by star-forming galaxies and/or small-scale intergalactic structure in the ionised IGM may dominate the reionisation budget, at least at reionisation's tail end (Davies et al. 2021; Cain et al. 2021; Cain, D'Aloisio, et al. 2024), perhaps requiring an increased photon budget to finish reionisation (Muñoz et al. 2024; Davies, Bosman, and Furlanetto 2024).

If the trend suggested by our fiducial measurement - that $\dot{N}_{\rm ion}$ increases with redshift - holds true to higher redshifts, then a $z \sim 5.3$ end to reionisation would likely require the entire process to be absorption-dominated. Indeed, this would be necessary in a scenario like that proposed by Muñoz et al. (2024), in which galaxies produce many more ionising photons that needed to re-ionise the universe by this time. This would be consistent with the recent measurements of the IGM clumping factor by Davies, Bosman, and Furlanetto (2024) - they find $C \sim 12$ at $z \leq 5$ and an upward trend towards $z = 6$. However, our measurement using the G23 data allows for scenarios in which $\dot{N}_{\rm ion}$ decreases at $z > 6$, in which case only the end stages may be dominated by recombinations.

### *4.3 Comparison to simulations*

There have been several recent attempts to reproduce Ly$\alpha$ forest and MFP measurements at $z \leq 6$ using numerical simulations of reionisation. These include semi-numerical approaches (e.g. Qin et al. 2024), post-processing RT (e.g. Cain et al. 2021), and RT coupled to hydrodynamics and galaxy formation (e.g. Kannan et al. 2022). In some cases, $\dot{N}_{\gamma}(z)$ is calibrated or fitted to reproduce these observations (e.g. Kulkarni et al. 2019; Cain et al. 2021; Qin et al. 2024), and in others it is predicted from an assumed galaxy model (e.g. Ocvirk et al. 2021; Lewis et al. 2022; Garaldi et al. 2022). The $\dot{N}_{\rm ion}$ in models that reproduce quasar observations sets a theoretical expectation that we can compare to our measurements.

We compare our fiducial measurement to $\dot{N}_{\rm ion}$ from numerical simulations in the top panel of Figure 3. The black solid curve and shaded region is our measurement and $1\sigma$ range, and the faded curves are results from several recent numerical simulations in the literature (referenced in the caption). All these agree reasonably well with the mean transmission of the Ly$\alpha$

j. This is accomplished by setting $\nu'' = \nu' = \nu$ in Equation (4-5).

forest and its large-scale fluctuations at $z < 6$. However, our fiducial measurement is a factor of $\sim 2$ above the simulations at $z \geq 5$, with most well below the $1\sigma$ range at all redshifts. At face value, this hints at a possible tension between simulations of re-ionisation's end and direct $\dot{N}_{\rm ion}$ measurements.

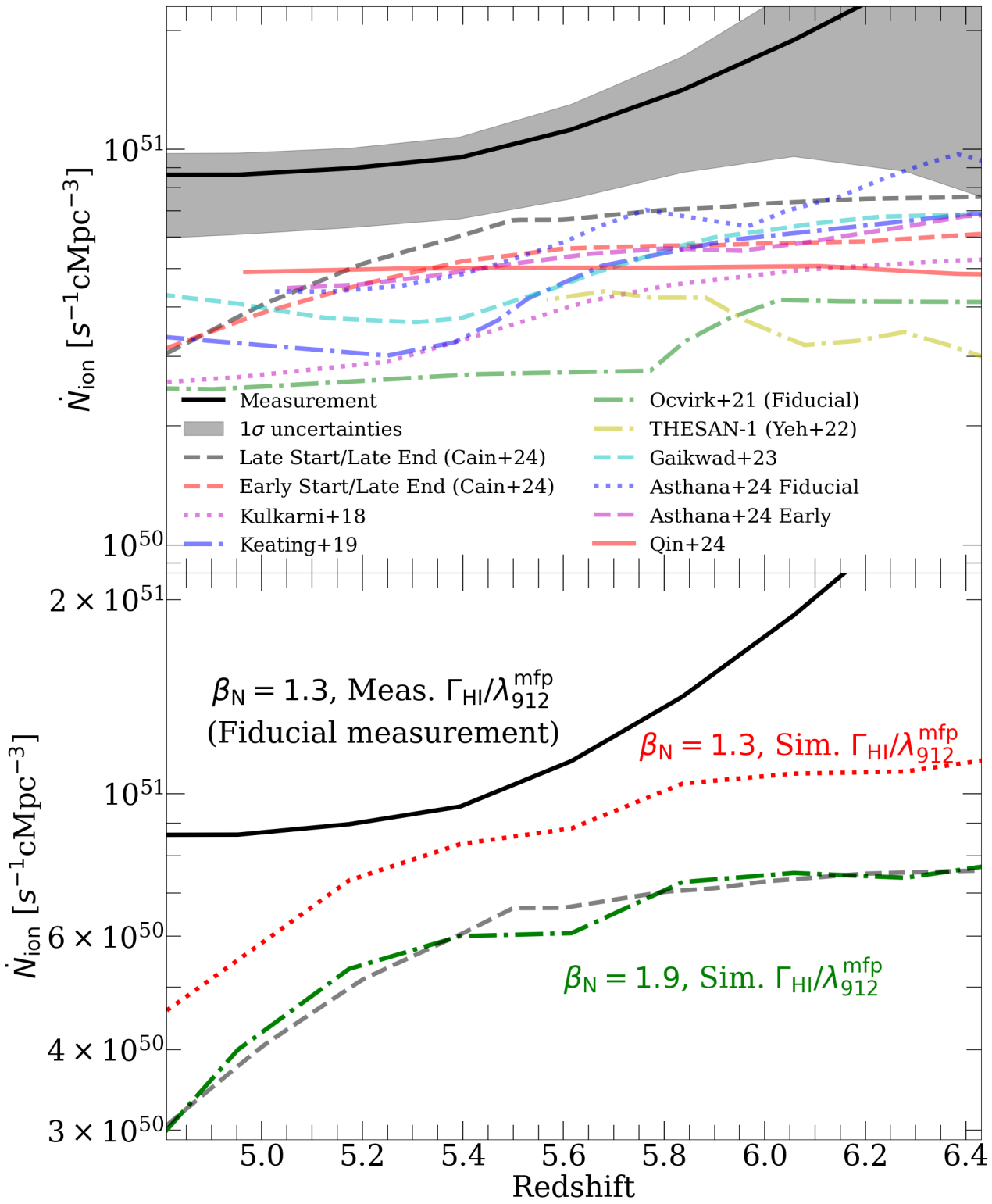

**Figure 3.** Comparison of our fiducial measurement of $\dot{N}_{\rm ion}$ to simulations that agree with the properties of the Ly$\alpha$ forest at $z < 6$. **Top**: our fiducial measurement, with $1\sigma$ uncertainties, compared with simulation results from Kulkarni et al. (2019), Keating, Weinberger, et al. (2020), Ocvirk et al. (2021), Yeh et al. (2023), Gaikwad et al. (2023), Asthana, Haehnelt, et al. (2024), Cain, Lopez, et al. (2024), and Qin et al. (2024). Our measurement is a factor of $\sim 2$ above most simulation results, suggesting a possible tension between measurements and simulations. **Bottom**: mock measurement of $\dot{N}_{\rm ion}$ using the $\Gamma_{\rm HI}$ and $\lambda_{912}^{\rm mfp}$ from the LATE START/LATE END model of Cain, Lopez, et al. (2024) (green dot-dashed) run with the FlexRT code compared to the simulation result (gray dashed). We assume $\beta_{\rm N} = 1.9$ and $\alpha = 1.5$, consistent with the IGM and source properties in the simulation. The agreement between these validates our formalism. The red dotted curve shows the same calculation assuming $\beta_{\rm N} = 1.3$, which lies above the simulation by a factor of $\sim 1.5$, potentially explaining some of the difference between the simulation and our measurement. We also show, for reference, our fiducial measurement (black solid, same as in the top panel), which assumes $\beta_{\rm N} = 1.3$ and the measured $\Gamma_{\rm HI}$ and $\lambda_{912}^{\rm mfp}$.

The bottom panel of Figure 3 explores the origin of this apparent disagreement. The gray-dashed curve shows (as in the top panel) $\dot{N}_{\rm ion}$ from the LATE START/LATE END model of Cain, Lopez, et al. (2024). We then take $\lambda_{912}^{\rm mfp}$ and $\Gamma_{\rm HI}$ from the simulation and repeat the procedure described in Section 2 to get a mock "measurement" of $\dot{N}_{\gamma}$, which is shown as the green dot-dashed curve. In this calculation, we assume $\beta_{\rm N} = 1.9$, consistent with the typical value of $\beta_{\rm N}$ seen for the ionising opacity model used in FlexRT (see Appendix B and Figure B2 of Cain, D'Aloisio, et al. (2024) for details). We also use $\alpha = 1.5$, the value used in the simulation. Our mock measurement agrees well with $\dot{N}_{\rm ion}$ from the simulation, validating the formalism used in this work. The red dotted curve is the same calculation, but assuming $\beta_{\rm N} = 1.3$, corresponding to a larger contribution to the IGM opacity from high column-density absorbers. The inferred $\dot{N}_{\rm ion}$ is a factor of $\sim 1.5$ higher than assuming $\beta_{\rm N} = 1.9$, which is a large fraction of the difference between simulations and our fiducial measurement. For reference, we also show the fiducial measurement from the top panel as the black solid curve. The difference between this and the red dotted curve arises from (10 – 20%-level) differences in $\Gamma_{\rm HI}$ and $\lambda_{912}^{\rm mfp}$ between the simulation and the measurement.

This comparison indicates that uncertainty in $\beta_{\rm N}$ may explain the difference between our measurement and $\dot{N}_{\rm ion}$ in simulations that reproduce the Ly$\alpha$ forest (see also Asthana, Kulkarni, et al. 2024). If $\beta_{\rm N}$ is close to 2, as it is in FlexRT, the IGM opacity is likely dominated by low column density, highly ionised absorbers (McQuinn, Oh, and Faucher-Giguère 2011). A value closer to 1 would indicate a large contribution from high column, self-shielding absorbers, and would demand a higher ionising output from galaxies. The true column density distribution at these redshifts is poorly understood. It likely is not well-described by a single power law, evolves in a complicated way during reionisation (Nasir et al. 2021), depends on the dynamics of small-scale structures that are challenging to resolve in reionisation simulations (Park et al. 2016; D'Aloisio et al. 2020; Chan et al. 2024; Gnedin 2024). The considerable uncertainty in measurements caused by $\beta_{\rm N}$ motivates further studies of the HI column density distribution.

### 4.4 Measurements of $\langle f_{\rm esc}\xi_{\rm ion}\rangle_{L_{\rm UV}}$

We can translate our $\dot{N}_{\rm ion}$ measurements into constraints on $\langle f_{\rm esc}\xi_{\rm ion}\rangle_{L_{\rm UV}}$ using Equation (1). Following B24, we compute $\rho_{\rm UV}(z)$ at $2.5 < z < 6$ using the measured UVLFs from Bouwens et al. (2021), and we use two limiting UV magnitudes, $M_{\rm UV} = -17$ and $-11$. In the top left panel of Figure 4, we show our constraints on $\langle f_{\rm esc}\xi_{\rm ion}\rangle_{L_{\rm UV}}$ using both cutoffs (black solid and magenta-dotted curves, respectively). Following B24, we assume that at $z > 4$, it is reasonable to treat $\dot{N}_{\rm ion}$ as dominated by the galaxy population. At $z < 4$, the ionising output of quasars likely begins to contribute significantly (Kulkarni, Worseck, and Hennawi 2019; Finkelstein and Bagley 2022; Smith et al. 2024) and may even dominate the ionising budget (Boutsia et al. 2021). As such, our estimates of $\langle f_{\rm esc}\xi_{\rm ion}\rangle_{L_{\rm UV}}$ must be interpreted as upper limits. At $z > 4$, the shaded regions show the $1\sigma$ uncertainties, including errors from the measurements of $\dot{N}_{\rm ion}$ and $\rho_{\rm UV}$. For consistency with B24, we estimate the latter from the reported errors on the amplitude of the UVLF in Bouwens et al. (2021). The black and magenta points show measurements from B24 assuming the same limiting values of $M_{\rm UV}$. At $z < 4$, we show only shaded regions denoting upper limits, as annotated in the figure.

At $z = 6$, the central values of our measurements are similar to those of B24, but our error bars are smaller. The agreement is coincidental, since our assumed $\Gamma_{HI}$ and $\lambda_{912}^{mfp}$ are higher than theirs at this redshift. At $z = 5$, our measurement is almost $1\sigma$ above that of B24, reflecting the difference between our $\dot{N}_{ion}$ and that of G23 seen in Figure 2. This arises in part from our more accurate treatment of RT effects, and also from a difference in assumed $\Gamma_{HI}$ and $\lambda_{912}^{mfp}$ measurements. Our $1\sigma$ upper limits are slightly higher than those of B24 at $z = 4$. At $z < 4$, our limits evolve little with redshift until $z = 2.5$, reflecting the lack of evolution in $\rho_{UV}$ and $\dot{N}_{ion}$. In reality, $\langle f_{esc}\,\xi_{ion}\rangle_{L_{UV}}$ probably continues to decline at $z < 4$ due to the increasing fraction of $\dot{N}_{ion}$ sourced by AGN. Note that we have not included any correction for the presence of neutral islands in Figure 4, since we have shown that this correction is small (and depends on the uncertain reionisation history).

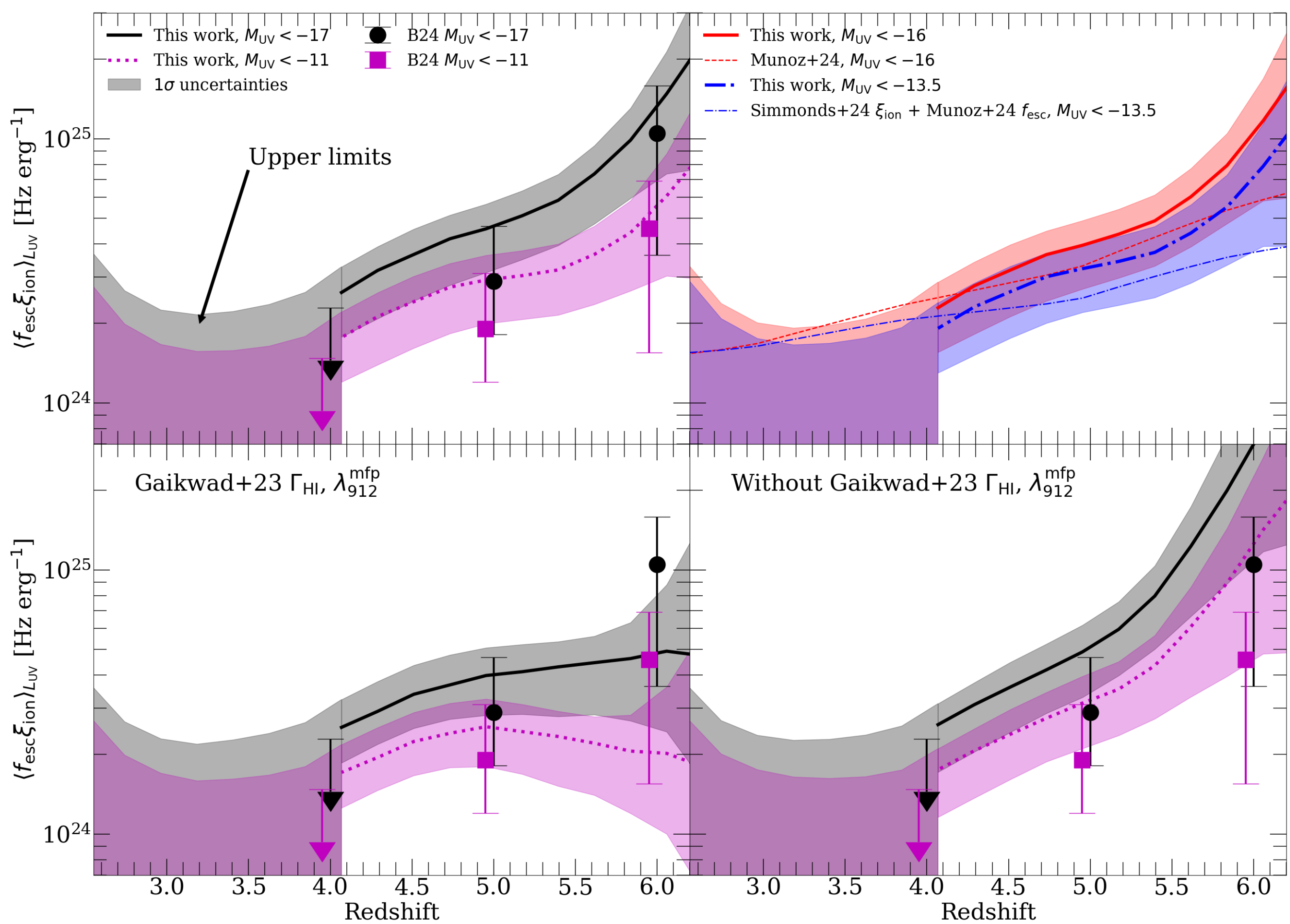

**Figure 4.** Constraints on $\langle f_{esc}\,\xi_{ion}\rangle_{L_{UV}}$ at $2.5 < z < 6$. **Top Left**: our fiducial constraints for $M_{UV} < -17$ (black solid curve) and $M_{UV} < -11$ (dotted magenta curve). The shaded regions at $z > 4$ indicate $1\sigma$ uncertainties, which include errors from both $\dot{N}_{ion}$ and $\rho_{UV}$ measurements. At $z < 4$, we treat our constraints as strict upper limits, since AGN likely dominate the ionising output of the source population at those redshifts. As such, we show shaded regions extending down to 0 at $z < 4$. The black and magenta points are constraints from B24. Our measurements are close to those of B24 at $z = 6$, nearly $1\sigma$ higher at $z = 5$, and at $z = 4$ our upper limit is slightly above theirs. **Top Right**: the red curves show that for $M_{UV} < -16$, our measurement roughly agrees with the model used in Muñoz et al. (2024), which uses $\xi_{ion}$ measurements from Simmonds et al. (2023) and the $f_{esc} - \beta_{UV}$ relation from Chisholm et al. (2022) (thin red dashed curve). The blue curves show the same comparison, but using the updated fit to measurements of $\xi_{ion}$ from the complete JADES sample in Simmonds et al. (2024). In this case, we find a similar level of agreement for $M_{UV} < -13.5$, which is more consistent with constraints on the faint-end cutoff of the UVLF (see text). **Bottom Left**: Same as in the top left panel, but measuring $\dot{N}_{ion}$ including only G23 measurements of $\Gamma_{HI}$ and $\lambda_{912}^{mfp}$ at $z \geq 5$. Here, $\langle f_{esc}\,\xi_{ion}\rangle_{L_{UV}}$ is nearly flat at $z > 5$ for $M_{UV} < -17$, declines slightly for $M_{UV} < -11$, and is well below the B24 measurements at $z = 6$. **Bottom Right:** like the bottom left, but *excluding* only the G23 data. In this case, $\langle f_{esc}\,\xi_{ion}\rangle_{L_{UV}}$ rises more steeply than in our fiducial measurement, and is more than $1\sigma$ *above* the $z = 6$ B24 measurements.

The top right panel compares our results to those predicted by the empirically-motivated model of Muñoz et al. (2024). They combined measurements of $\xi_{ion}$ from Simmonds et al. (2023)[k] with the $f_{esc} - \beta_{UV}$ relation calibrated by Chisholm et al. (2022) to estimate $\dot{N}_{ion}$ during reionisation. Both the $\xi_{ion}$ measurements by Simmonds et al. (2023) and $f_{esc}$ predicted by the $\beta_{UV}$–$f_{esc}$ increase with $M_{UV}$ (that is, for fainter galaxies). As such, the prediction for $\langle f_{esc}\xi_{ion}\rangle_{L_{UV}}$ of the Muñoz et al. (2024) model increases for fainter $M_{UV}$ cutoffs, which is the opposite of the dependence of our measurements. Given this, we can ask what cutoff produces the best agreement between their model and our measurement. The red solid and thin dashed curves show that our measurement agrees best

k. See also Endsley et al. (2024), Prieto-Lyon et al. (2023), Pahl et al. (2024), Atek et al. (2024), and Meyer et al. (2024) for other observational estimates of $\xi_{ion}$.

with their model for $M_{\rm UV} < -16$. This agreement echos the conclusion in their work that relatively bright UV cutoffs are required to bring their model into agreement with observations supporting a late end to reionisation.

The recent re-measurements of $\xi_{\rm ion}$ presented in Simmonds et al. (2024) predict significantly lower $\xi_{\rm ion}$, and weaker dependence on $M_{\rm UV}$, than found in Simmonds et al. (2023). Updating the $\xi_{\rm ion}$ used in the Muñoz et al. (2024) model with their new result and assuming $M_{\rm UV} < -13.5$ gives the thin blue dashed curve, which agrees reasonably well with our measurement assuming the same cutoff (solid blue). This indicates that the updated $\xi_{\rm ion}$ measurements from Simmonds et al. (2024) relieve some of the apparent tension between the Muñoz et al. (2024) model and a late end to reionisation. This cutoff is fainter than currently probed by lensed galaxies (Atek et al. 2018), and as such does not violate existing constraints on the faint-end turnover of the UVLF at these redshifts.

At $z > 5$, our measurements imply steeper redshift evolution of $\langle f_{\rm esc}\xi_{\rm ion}\rangle_{L_{\rm UV}}$ than in the Muñoz et al. (2024) model. This is a result of the rapid decline in the MFP at $z > 5$, which pushes $\dot{N}_{\rm ion}$ up quickly approaching $z = 6$. We measure a factor of $\approx 2$ increase in $\dot{N}_{\rm ion}$ between $z = 5$ and 6, which translates into a similar increase in $\langle f_{\rm esc}\xi_{\rm ion}\rangle_{L_{\rm UV}}$, since $\rho_{\rm UV}$ stays roughly constant. This evolution is qualitatively similar to that required in several prior works that match the evolution of the observed mean transmission of the Ly$\alpha$ forest (Kulkarni et al. (2019), Keating, Kulkarni, et al. (2020), and Ocvirk et al. (2021), see also Cain, D'Aloisio, et al. (2024)).

In the bottom panels, we show how our results change when we use different sub-sets of $\Gamma_{\rm HI}$ and $\lambda_{912}^{\rm mfp}$ measurements. The bottom left panel is the same as the top left, but using only the G23 measurements, while the bottom right excludes them. In the former case, $\langle f_{\rm esc}\xi_{\rm ion}\rangle_{L_{\rm UV}}$ increases by a factor of $\approx 2$ from $z = 2.5$ to 5, then plateaus at $z > 5$ for $M_{\rm UV} < -17$ (and declines slightly for $M_{\rm UV} < -11$). In the latter, $\langle f_{\rm esc}\xi_{\rm ion}\rangle_{L_{\rm UV}}$ rises more steeply than in our fiducial measurement, eclipsing the $z = 6$ measurements from B24 by $\approx 2\sigma$. In this scenario, $\langle f_{\rm esc}\xi_{\rm ion}\rangle_{L_{\rm UV}}$ increases by a factor of $\sim 3-4$ (depending on the $M_{\rm UV}$ cutoff) between $z = 5$ and 6. These different sets of measurements have very different implications for the ionising properties of $z > 5$ galaxies, reflect the large uncertainty in our knowledge of galaxy ionising properties at these redshifts. Thus, converging on precise high-$z$ measurements of $\Gamma_{\rm HI}$ and $\lambda$ is critical for understanding the ionising output of galaxies as reionisation is ending.

### 4.5 Measurements of $f_{\rm esc}$

Lastly, we turn to perhaps the most uncertain parameter in reionisation models, $f_{\rm esc}$. We can constrain the population averaged $f_{\rm esc}$ by taking advantage of the redshift and $M_{\rm UV}$-dependent fits to measurements of $\xi_{\rm ion}$ provided in Simmonds et al. (2023) and Simmonds et al. (2024). We can combine measurements of the UVLF with this result to measure the *intrinsic* ionising photon production rate averaged over the galaxy population,

$$\dot{N}_{\rm ion}^{\rm intr.} = \int_{-\infty}^{M_{\rm UV}^{\rm cut}} dM_{\rm UV}\frac{dn}{dM_{\rm UV}}L_{\rm UV}\xi_{\rm ion}(z, M_{\rm UV}) \tag{12}$$

where $\frac{dn}{dM_{\rm UV}}$ is the UVLF. Then the population-averaged escape fraction is

$$\langle f_{\rm esc}\rangle_{\dot{n}_\gamma^{\rm intr.}} = \frac{\dot{N}_{\rm ion}}{\dot{N}_{\rm ion}^{\rm intr.}} \tag{13}$$

The sub-script $\dot{n}_\gamma^{\rm intr.}$ indicates that the average is weighted by the intrinsic ionising output of individual galaxies. A caveat is that the measured ionising efficiency, which we call $\xi_{\rm ion}^0$, is related to the true one by $\xi_{\rm ion}^0 = \xi_{\rm ion}(1 - f_{\rm esc})$, since escaping ionising radiation does not contribute to the recombination emission used to measure $\xi_{\rm ion}$. We correct for this approximately by assuming that $\xi_{\rm ion} = \xi_{\rm ion}^0/(1 - \langle f_{\rm esc}\rangle_{\dot{n}_\gamma^{\rm intr.}})$, which ignores any dependence of $f_{\rm esc}$ on $M_{\rm UV}$. Under this approximation, it is straightforward to show that $f_{\rm esc} = f_{\rm esc}^0/(1 + f_{\rm esc}^0)$, where $f_{\rm esc}^0$ is Equation (13) evaluated assuming $\xi_{\rm ion} = \xi_{\rm ion}^0$. Since Equation (12) uses measured galaxy properties to get $\dot{N}_{\rm ion}^{\rm intr.}$, the $\dot{N}_{\rm ion}$ that is applied in Equation 13 should be the contribution from galaxies only, without the AGN contribution. To avoid complications and uncertainties associated with quantifying the AGN contribution, we instead use our measured $\dot{N}_{\rm ion}$ and interpret our $z < 4$ results as upper limits, as we do in Figure 4.

We show our fiducial results for this quantity in the top left panel of Figure 5 for $M_{\rm UV} < -17$ and $-11$ (thick curves). These measurements use the updated $\xi_{\rm ion}$ estimates from Simmonds et al. (2024). In the top-right panel, we show the same results assuming the prior findings of Simmonds et al. (2023). The black point shows the $f_{\rm esc} = 0.085$ measured at $z \sim 3$ by Pahl et al. (2021) for their sample of faint ($L < L_*$) galaxies, which are more likely to be analogous to the galaxies that dominated the ionising photon budget during reionisation (Atek et al. 2024, although see arguments to the contrary in Naidu et al. 2022; Matthee et al. 2022). We include uncertainties in the power-law intercept reported in the $\xi_{\rm ion}$ fits in Simmonds et al. (2023) and Simmonds et al. (2024) in our error budget[1] on $\langle f_{\rm esc}\rangle_{\dot{n}_\gamma^{\rm intr.}}$, along with the errors on $\dot{N}_{\rm ion}$ and $\rho_{\rm UV}$ already included in Figure 4.

The redshift evolution we infer for $\langle f_{\rm esc}\rangle_{\dot{n}_\gamma^{\rm intr.}}$ is qualitatively similar to that of $\langle f_{\rm esc}\xi_{\rm ion}\rangle_{L_{\rm UV}}$. At $z < 4$ and $M_{\rm UV} < -17$, our upper limit is not much higher than the Pahl et al. (2021) measurement of 8.5% at $z = 3$, and is actually slightly below this value for $M_{\rm UV} < -11$. Between $z = 4$ and 5, our measurements increase by a factor of $\approx 2$, and by another factor of $\approx 2$ from $z = 5$ to 6. This steep increase occurs because $\rho_{\rm UV}$ decreases between $z = 4$ and 6 and $\xi_{\rm ion}$ remains flat, such that an increasing $f_{\rm esc}$ is required to explain the increase in $\dot{N}_{\rm ion}$ over

1. The size of our error bars are likely an under-estimate, since we neglect uncertainty in the $M_{\rm UV}$ and redshift dependence of the $\xi_{\rm ion}$ measurements. The same is true of our errors propagated from $\rho_{\rm UV}$, since we do not account for uncertainty in the shape of the UVLF - only its amplitude (following Bosman and Davies (2024)).

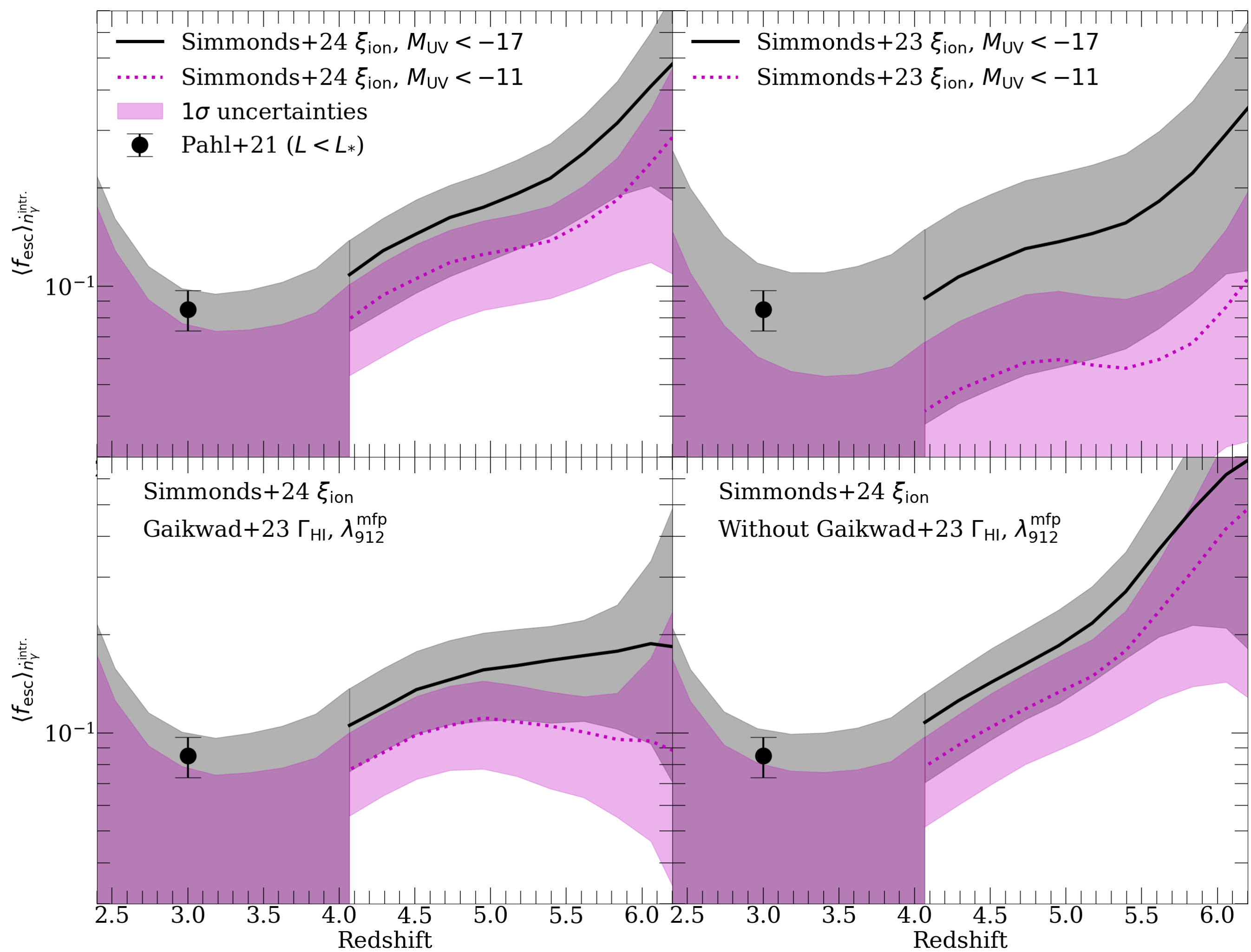

**Figure 5.** Constraints on the $\dot{n}_\gamma^{\rm intr.}$-weighted average escape $f_{\rm esc}$ of the galaxy population at $2.5 < z < 6$. **Top left**: Fiducial constraints for $M_{\rm UV} < -17$ and $-11$. At $z > 4$ we report our constraints as measurements, and as upper limits at $z < 4$ (as in Figure 4). The black point shows $f_{\rm esc} = 0.085$ measured by Pahl et al. (2021) at $z \sim 3$ for faint ($L < L_*$) galaxies. **Top right**: the same, but using the prior results for $\xi_{\rm ion}$ from Simmonds et al. (2023). **Bottom Left:** same as the top left, but using only $\Gamma_{\rm HI}$ and $\lambda_{912}^{\rm mfp}$ measurements from G23 at $z > 5$. **Bottom Right:** the same, but excluding the G23 results. See text for details.

this range. Note that because the AGN contribution to $\dot{N}_{\rm ion}$ grows between $z = 4$ and 2.5, $\langle f_{\rm esc} \rangle_{\dot{n}_\gamma^{\rm intr.}}$ likely declines with cosmic time over this range.

The top right panel shows the same measurements, but using the older results for $\xi_{\rm ion}$ from Simmonds et al. (2023). Note that the error bars are much larger because of the higher uncertainties on $\xi_{\rm ion}$ in that work. We find systematically lower $\langle f_{\rm esc} \rangle_{\dot{n}_\gamma^{\rm intr.}}$, reflecting the higher $\xi_{\rm ion}$ values measured in that work. We also see noticeably different redshift evolution, especially for the $M_{\rm UV} < -11$ case. The evolution of $\langle f_{\rm esc} \rangle_{\dot{n}_\gamma^{\rm intr.}}$ is significantly flatter, and does not reach 10% until $z = 6$ in that case. The dependence on the assumed $M_{\rm UV}$ cutoff is also much stronger, with a factor of $\approx 2$ difference between $M_{\rm UV} < -11$ and $M_{\rm UV} < -17$, compared to a 15–20% difference in the upper left. This arises from the fact that in the Simmonds et al. (2023) results, fainter galaxies have much higher $\xi_{\rm ion}$ than bright galaxies. In the updated measurements, $\xi_{\rm ion}$ does not depend strongly on $M_{\rm UV}$. This finding further highlights the much lower ionising output expected for galaxies based on the new Simmonds et al. (2024), which helps relieve the "too many photons" problem described in Muñoz et al. (2024).

The bottom left is the same as the top left, but using only the G23 measurements at $z > 5$ to measure $\dot{N}_{\rm ion}$. Consistent with Figure 4, we find a more modest evolution in $\langle f_{\rm esc} \rangle_{\dot{n}_\gamma^{\rm intr.}}$ than our fiducial result. We find $\langle f_{\rm esc} \rangle_{\dot{n}_\gamma^{\rm intr.}} < 15\%$ at $z = 2.5$ and 4, $\approx 15\%$ at $z = 5$, and $\approx 19\%$ at $z = 6$ for $M_{\rm UV} < -17$. These numbers become 12%, 11%, and 9% for $M_{\rm UV} < -11$. Within the uncertainties, these results are consistent with no evolution in $\langle f_{\rm esc} \rangle_{\dot{n}_\gamma^{\rm intr.}}$ at $2.5 < z < 6$. In the bottom right, we show the result without G23 measurements included. We find steep evolution in $\langle f_{\rm esc} \rangle_{\dot{n}_\gamma^{\rm intr.}}$ at $z > 4$, rising to 40 – 60% by $z = 6$ (depending on the cutoff $M_{\rm UV}$)[m].

m. As noted in Section 4.3, a higher value of $\beta_{\rm N}$ would result in lower $\dot{N}_{\rm ion}$ and correspondingly lower $f_{\rm esc}$ measurements, possibly by a factor of 1.5 or more (see Figure 3). Uncertainty from $\beta_{\rm N}$ is reflected in the error bars in Our upper limits are consistent with population-averaged escape fractions of $\lesssim 10\%$ at $z \leq 4$. This agrees with recent efforts to measure $f_{\rm esc}$ directly at these redshifts (e.g. Smith et al. 2018; Smith et al. 2020; Pahl et al. 2021; Kerutt et al. 2024) and with indirect determinations based on simulations (Finkelstein et al. 2019; Yeh et al. 2023; Choustikov et al. 2024). Indeed, escape fractions at these redshifts may be well below 10% if the ionising background is sustained mainly by the quasar population (Boutsia et al. 2021). In this case, a self-consistent measurement of galaxy ionising properties at $z < 4$ should also take into a account the luminosity function and ionising properties of the quasar population. Such an endeavour could be aided by using the properties of the HeII Ly$\alpha$ forest to distinguish between galaxy and quasar ionising output (e.g. McQuinn and Worseck 2014; D'Aloisio et al. 2017; Gaikwad, Davies, and Haehnelt 2025). We defer such an investigation to future work.

At $z > 4$, when $f_{\rm esc}$ becomes impossible to measure directly, we find a factor of $\sim 1.5-2$ increase between $z = 4$ and 5 in our fiducial measurement. Indeed, the actual evolution is probably steeper than this, because AGN contribute more to $\dot{N}_{\rm ion}$ at $z = 4$ than at $z = 5$. This finding suggests modest but significant evolution in galaxy ionising properties, perhaps due to evolution in the properties of the ISM/CGM between $z = 4$ and 5 (e.g. Kakiichi and Gronke 2021; Kimm et al. 2022). At $z > 5$, our results are sensitive to the choice of $\Gamma_{\rm HI}$ and $\lambda_{912}^{\rm mfp}$ measurements used. Using the only indirect measurements of G23 suggests flat evolution in $f_{\rm esc}$ at $z > 5$, while ignoring these measurements gives a factor of $\sim 3$ increase between $z = 5$ and 6. These findings motivate further efforts to reduce uncertainty on measurements of $\Gamma_{\rm HI}$ and $\lambda_{912}^{\rm mfp}$ at $z > 5$, and to understand why different measurement methods give significantly different results.

### 4.6 Comparison to indirect determinations of $f_{\rm esc}$

We briefly compare our $f_{\rm esc}$ results to several indirect observational and theoretical determinations. These include empirically motivated estimates of $f_{\rm esc}$ based on measurements at lower redshift and estimates from numerical simulations. In Figure 6, we show the comparison to our maximum likelihood measurements of $\langle f_{\rm esc} \rangle_{\dot{n}_\gamma^{\rm intr.}}$ for all three sets of $\Gamma_{\rm HI}$ and $\lambda_{912}^{\rm mfp}$ measurements with $M_{\rm UV} < -11$. The thin dot-dashed green curve shows the empirical model of Muñoz et al. (2024) (for $M_{\rm UV} < -11$), based on the $\beta_{\rm UV}$-$f_{\rm esc}$ relation from Chisholm et al. (2022), and the $\beta_{\rm UV} - M_{\rm UV}$ relation from Zhao and Furlanetto (2024). Unlike Muñoz et al. (2024), however, we use the updated $\xi_{\rm ion}$ measurements from Simmonds et al. (2024). We also show the population-averaged $f_{\rm esc}$ from THESAN (red dashed, Yeh et al. 2023) and SPHINX (blue dotted, Rosdahl et al. 2022), and the global $f_{\rm esc}$ from Finkelstein et al. (2019) for all galaxies (black dot-dashed) and faint ($M_{\rm UV} > -15$) galaxies (cyan dot-dashed). Note that we only show these down to $z = 4$, since our constraints at $z < 4$ are upper limits.

Figure 5.

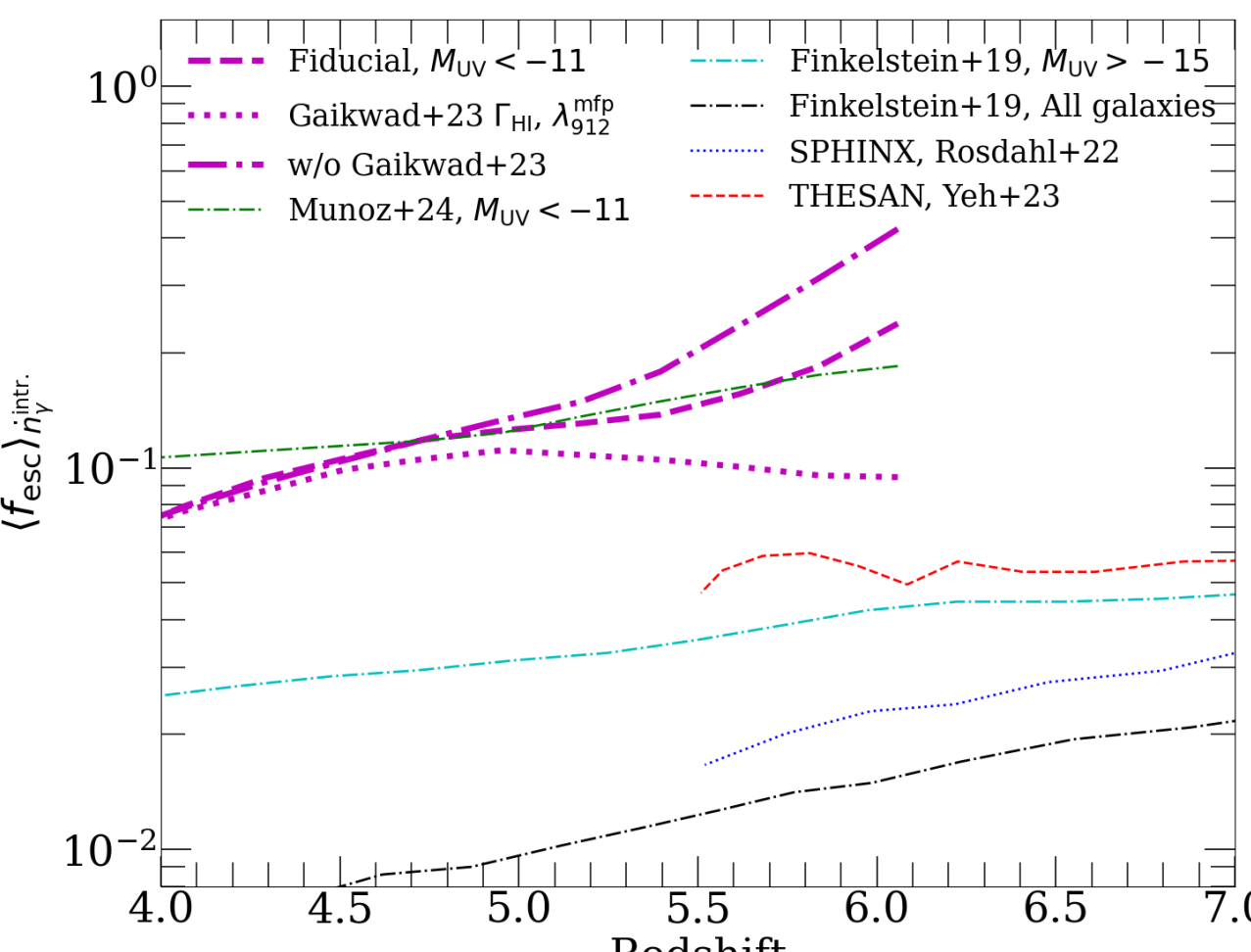

**Figure 6.** Comparison of our $\langle f_{\rm esc} \rangle_{\dot{n}_\gamma^{\rm intr.}}$ measurements to several indirect determinations. The thin green dot-dashed curve shows the model used in Muñoz et al. (2024) based off the $f_{\rm esc}$-$\beta_{\rm UV}$ relation calibrated by Chisholm et al. (2022) at lower redshifts, and using the latest $\xi_{\rm ion}$ measurements from Simmonds et al. (2024). The other curves show simulation results from Finkelstein et al. (2019), Rosdahl et al. (2022), and Yeh et al. (2023). See text for discussion.

All the simulation results lie below the measurements to varying degrees at $z > 4$, although their redshift evolution is comparable to our fiducial result. The disagreement is greatest for the measurement without the G23 data, which implies steeper redshift evolution than in the $f_{\rm esc}$–$\beta_{\rm UV}$ model, or any of the simulation curves. Note that using a brighter cutoff $M_{\rm UV}$ increases $f_{\rm esc}$, worsening this disagreement. At face-value, this suggests that simulations may be under-estimating galaxy escape fractions. Another possibility is that AGN contribute significantly to $\dot{N}_{\rm ion}$ at these redshifts (Madau et al. (2024), Smith et al. (2024), and Dayal et al. (2024), although see Jiang et al. (2025)), or current measurements of $\xi_{\rm ion}$ are underestimates. Both scenarios would reduce our measured $f_{\rm esc}$.

The empirical $f_{\rm esc} - \beta_{\rm UV}$ relation agrees reasonably well with our measurement for $M_{\rm UV} < -11$, predicting $f_{\rm esc} \sim 10\%$ at $z \lesssim 5$ (see also upper right panel of Figure 4). At $5 < z < 6$ it agrees best with the fiducial measurement, while the measurements only with (without) data from G23 at $z > 5$ fall slightly below (above) the model. One caveat to this agreement is that the measured $f_{\rm esc}$ becomes significantly higher for $M_{\rm UV} < -17$ (Figure 5), while the same is not true of the Muñoz et al. (2024) model. As such, our fiducial measurement would lie above the prediction of the $\beta_{\rm UV}$-$f_{\rm esc}$ relation at $z > 5$ if the UVLF cuts off significantly brighter than $M_{\rm UV} = -11$.

Recently, Qin et al. (2024) used semi-numerical simulations of reionisation to constrain both $\dot{N}_{\rm ion}$ (red solid curve in Figure 3) and the dependence of the escape fraction on both redshift and galaxy properties. Their constraints included UVLF measurements, the CMB optical depth, and Ly$\alpha$ forest data, although they found the latter to be by far the most constraining on galaxy ionising properties. They found that the escape fraction is much higher in lower-mass halos, with a normalisation that increases with decreasing redshift, $f_{\rm esc} \propto (1+z)^{-1.6}$ (note that they do not plot $\langle f_{\rm esc} \rangle_{\dot{n}_{\gamma}^{\rm intr.}}$, so we cannot show it on Figure 6). The latter is the opposite the trend we find – that $f_{\rm esc}$ grows towards higher redshifts, especially from $z = 5 - 6$ (although the strong dependence on halo mass may make the $\dot{n}_{\rm intr.}$-weighted average evolve less quickly). This is in part due to the fact that our $\dot{N}_{\rm ion}$ increases rapidly with redshift across this redshift range while theirs remains flat (Figure 3). It is also unclear whether the $\xi_{\rm ion}$ assumed in their model matches the parameterization we use here based on JWST measurements, which could further explain the difference between our results.

## 5. Conclusions

In this work, we have provided new measurements of the ionising emissivity of the galaxy population at $2.5 < z < 6$. Our measurements take into account the effects of RT without using the local source approximation. We also developed a formalism to account for the possible presence of neutral islands at $z > 5$, when reionisation may be ongoing. We present measurements of the global ionising emissivity, the average escaping ionising efficiency, and the average escape fraction of the high-redshift galaxy population. Our main conclusions are summarised below:

- We measure $\dot{N}_{\rm ion} = 10.04^{+2.72}_{-4.66}$, $7.05^{+1.41}_{-2.17}$, $8.70^{+1.15}_{-2.55}$, and $17.67^{+5.62}_{-8.29} \times 10^{50}$ s$^{-1}$ cMpc$^{-3}$ at $z = 2.5$, 4, 5, and 6, respectively, at $1\sigma$ confidence. Our measurement of $\dot{N}_{\rm ion}$ at $z = 6$ is nearly $2\sigma$ higher than that of G23, largely due to the shorter $\lambda_{912}^{\rm mfp}$ assumed here. At $z = 5$, our measurement is higher than theirs by $\approx 0.8\sigma$. The latter is in large part due to our inclusion of RT effects. At $z \leq 4.5$, our measurements are consistent with those of Becker and Bolton (2013). We find that including the opacity from neutral islands at a level consistent with expectations from reionisation simulations has a $\lesssim 15\%$ effect on these measurements. The change is modest because the effect of islands on $\dot{N}_{\rm abs}$ and $\lambda$ approximately cancels in Equation (6). This suggests that the standard approach to measuring $\dot{N}_{\rm ion}$ may work even while reionisation is ongoing.
- We measure the $5.3 < z < 6$ photon budget, likely covering the last 10 – 20% of reionisation, to be $N_{\gamma/H}^{z=5.3-6} = 1.10^{+0.21}_{-0.39}$. This is a factor of $\approx 4$ higher than the budget required to complete the last 20% of reionisation in the absence of absorptions in the ionised IGM. This elevated photon budget is reflected in our measurements of $f_{\rm esc}$ at $z > 5$, which are a factor of $\gtrsim 2$ higher than predicted by simulations. Our findings suggest that at least the tail-end of reionisation may have been absorption-dominated.
- We find a factor $\sim 2$ tension between our fiducial measurement of $\dot{N}_{\rm ion}$ and the emissivity required in reionisation simulations that reproduce the observed transmission of the $z < 6$ Ly$\alpha$ forest. A significant fraction (roughly half) of that difference may arise from differences in the shape of the HI column density distribution in the simulations compared to that assumed in the measurement.
- We constrain the collective escaping ionising efficiency of the galaxy population, assuming the UVLF from Bouwens et al. (2021), to be $\log_{10}(\langle f_{\rm esc} \xi_{\rm ion} \rangle_{L_{\rm UV}} / [{\rm erg^{-1} Hz}]) < 24.61$, $< 24.49$, $= 24.67^{+0.10}_{-0.17}$ and $= 25.12^{+0.15}_{-0.28}$ at $z = 2.5$, 4, 5, and 6, respectively, for $M_{\rm UV} < -17$ and at $1\sigma$ confidence. These numbers become $< 24.49$, $< 24.32$, $= 24.47^{+0.10}_{-0.17}$, and $= 24.75^{+0.15}_{-0.28}$, respectively, for $M_{\rm UV} < -11$. Our measurement of $\langle f_{\rm esc} \xi_{\rm ion} \rangle_{L_{\rm UV}}$ is similar to that of B24 at $z = 6$, but is almost $1\sigma$ higher at $z = 5$. At $z = 4$, our upper limit is slightly higher than theirs. We find that our upper limits on $\langle f_{\rm esc} \xi_{\rm ion} \rangle_{L_{\rm UV}}$ remain roughly flat at $2.5 < z < 4$, but we find a factor of 1.5 – 2 increase in escaping efficiency between $z = 4$ and 5, and another factor of 2 between $z = 5$ and 6. The evolution between $z = 5$ and 6 owes to the fact that the measured $\lambda_{912}^{\rm mfp}$ declines at those redshifts more quickly than does $\Gamma_{\rm HI}$.
- We constrain the population-averaged ionising escape fraction $\langle f_{\rm esc} \rangle_{\dot{n}_{\gamma}^{\rm intr.}}$, leveraging the redshift and $M_{\rm UV}$-dependent fit to measurements of $\xi_{\rm ion}$ provided by Simmonds et al. (2023) and Simmonds et al. (2024). For $M_{\rm UV} < -17$, we measure $\langle f_{\rm esc} \rangle_{\dot{n}_{\gamma}^{\rm intr.}} < 0.172$, $< 0.131$, $= 0.178^{+0.048}_{-0.058}$, and $= 0.385^{+0.168}_{-0.186}$ at $z = 2.5$, 4, 5, and 6, respectively, at $1\sigma$ confidence. These numbers become $\langle f_{\rm esc} \rangle_{\dot{n}_{\gamma}^{\rm intr.}} < 0.138$, $< 0.096$, $= 0.126^{+0.034}_{-0.041}$, and $= 0.224^{+0.098}_{-0.108}$ for $M_{\rm UV} < -11$. At $z \lesssim 4$, our upper limits are consistent with measurements of $f_{\rm esc}$ in galaxies at those redshifts suggesting values $\lesssim 10\%$. However, we find a factor of 2 – 3 increase between $z = 4$ and 6, suggesting evolution in the ionising properties of the galaxy population at these redshifts.
- We repeated our measurements using two sub-sets of $\Gamma_{\rm HI}$ and $\lambda_{912}^{\rm mfp}$ measurements. The first excludes all measurements at $z > 5$ except those of G23, and the second set includes all measurements except those of G23. The evolution of $\Gamma_{\rm HI}$ ($\lambda_{912}^{\rm mfp}$) with redshift is steeper (shallower) with redshift in the G23-only case, resulting in much flatter $\dot{N}_{\rm ion}$ evolution. Conversely, leaving out the G23 measurements results in a steeper evolution in $\dot{N}_{\rm ion}$. These differences translate into very different estimates of the redshift evolution of $f_{\rm esc}$ at $z > 5$ – nearly flat in the former case, and a factor of 4 increase to $\approx 60\%$ (40%), for $M_{\rm UV} < -17$ ($-11$), in the latter.

We find that RT has a modest, but significant effect on measurements of the ionising output of galaxies at $4 < z < 6$. They also highlight uncertainties in measurements of IGM conditions at $z > 5$, which translate into large uncertainties in the evolution of galaxies ionising properties near reionisation's end. Perhaps most notably, the implied evolution of $f_{\rm esc}$ varies considerably at $z > 5$ when different $\Gamma_{\rm HI}$ and $\lambda_{912}^{\rm mfp}$ data sets are used to calculate $\dot{N}_{\rm ion}$. These findings motivate further efforts to pin down these measurements at $z > 5$.

**Acknowledgments** The authors thank Sarah Bosman, Simeon Bird, and Rolf Jansen for helpful discussions. We also thank George Becker, Frederick Davies, Steven Finkelstein, Brent Smith, Yongda Zhu, and Rogier Windhorst for helpful com-ments on the draft version of this manuscript.

**Funding Statement** CC was supported by the Beus Center for Cosmic Foundations. AD was supported by grants NSF AST-2045600 and JWSTAR02608.001-A. JBM was supported by NSF Grants AST-2307354 and AST-2408637, and by the NSF-Simons AI Institute for Cosmic Origins.

**Competing Interests** The authors are not aware of any competing interests connected to this work.

**Data Availability Statement** The data underlying this article will be shared upon reasonable request to the corresponding author.

**Ethical Standards** The research meets all ethical guidelines, including adherence to the legal requirements of the study country.

## Appendix 1. Connection to the standard radiative transfer formalism

In this appendix, we show that our RT formulation (Equation 3-5) is equivalent to the standard formalism presented e.g. in Haardt and Madau (1996). The angle-averaged specific intensity at redshift $z$ and frequency $\nu$ is

$$J_\nu(z) = \frac{c}{4\pi}\int_z^\infty \frac{dz'}{(1+z')H(z')}\frac{(1+z)^3}{(1+z')^3}\epsilon_{\nu'}(\nu',z')e^{-\tau_{\rm eff}(\nu,z,z')}, \tag{14}$$

where primed and un-primed variables have the same meaning as in Section 2. Here, $\epsilon_{\nu'}(\nu', z')$ is the proper energy emissivity of sources per unit $\nu'$, and $\tau_{\rm eff}$ is the effective optical depth between $z'$ and $z$, given by (see e.g. Equation 8 of McQuinn 2016),

$$\tau_{\rm eff}(\nu,z,z') = \int_{x(z')}^{x(z)}\frac{dx}{1+z(x)}\lambda_{\nu''}^{-1} = \int_{t'}^{t}dt'' c\kappa_{\nu''}, \tag{15}$$

where the first integral is over co-moving distance $x$.[n] Note that $\nu'' = \nu\frac{1+z''}{1+z}$, analogously to the singly primed variable. In the second equality, we have used the fact that $cdt = \frac{dx}{1+z(x)}$ and $\kappa_\nu \equiv \lambda_\nu^{-1}$, which recovers the expression in the exponential of Equation (5). Finally, the ionisation rate is given by (Equation 10 of Haardt and Madau (1996))

$$\Gamma_{\rm HI} = 4\pi\int d\nu\frac{J_\nu}{h_{\rm p}\nu}\sigma_{\rm HI}^{\nu} \tag{17}$$

To show that Equation (14-17) are equivalent to Equation (3-5), we first change variables to $t$ in Equation (14), recognising

n. Often, the LyC opacity of the IGM is modelled as arising from a population of Poisson distributed absorbers, in which case the effective optical depth takes the less general form

$$\tau_{\rm eff}(\nu,z,z') = \int_z^{z'}dz''\int_0^\infty dN_{\rm HI}\frac{\partial^2 N}{\partial N_{\rm HI}\partial z''}(1-e^{-\tau}), \tag{16}$$

where $\frac{\partial^2 N}{\partial N_{\rm HI}\partial z}$ is the HI column density distribution and $\tau = N_{\rm HI}\sigma_{\rm HI}^{\nu'}$.

that $dt = -\frac{dz}{(1+z)H(z)}$, and substitute Equation (15) into Equation (14), which gives

$$J_\nu(z) = \frac{c(1+z)^3}{4\pi} \int_0^t dt' \frac{\epsilon_{\nu'}(\nu', t')}{(1+z')^3} e^{-\int_{t'}^{t} dt'' c\kappa_{\nu''}} \tag{18}$$

The proper emissivity $\epsilon_{\nu'}(\nu', t')$ can be re-written in terms of the co-moving photon emissivity per unit observed frequency $\nu'$ (see Equation (4)) as

$$\epsilon_{\nu'}(\nu', t') = (1+z')^3 \frac{1+z}{1+z'} h_p \nu' \dot{N}^{\nu}_{\rm ion}(t', \nu') \tag{19}$$

where the factor of $\frac{1+z'}{1+z}$ arises from the fact that $\dot{N}^{\nu}_{\rm ion}$ is a derivative with respect to co-moving frequency $\nu$ rather than $\nu'$. Putting this into Equation (18) gives

$$J_\nu(z) = \frac{c(1+z)^3}{4\pi} \int_0^t dt' \frac{1+z}{1+z'} h_p \nu' \dot{N}^{\nu}_{\rm ion}(t', \nu') e^{-\int_{t'}^{t} dt'' c\kappa_{\nu''}} \tag{20}$$

Recognising that $\nu = \frac{1+z}{1+z'}\nu'$ and putting Equation (20) into Equation (17) yields

$$\Gamma_{\rm HI} = 4\pi \int \frac{d\nu}{h_p \nu} \sigma^{\nu}_{\rm HI} \frac{c(1+z)^3}{4\pi} \int_0^t dt' h_p \nu \dot{N}^{\nu}_{\rm ion}(t', \nu') e^{-\int_{t'}^{t} dt'' c\kappa_{\nu''}} \tag{21}$$

which simplifies to

$$\Gamma_{\rm HI} = c(1+z)^3 \int d\nu \sigma^{\nu}_{\rm HI} \int_0^t dt' \dot{N}^{\nu}_{\rm ion}(t', \nu') e^{-\int_{t'}^{t} dt'' c\kappa_{\nu''}} \tag{22}$$

Finally, we can multiply inside the integral over $\nu$ by $\lambda_\nu(t)\kappa_\nu(t) = 1$ and group terms to obtain

$$\Gamma_{\rm HI} = (1+z)^3 \int d\nu \lambda_\nu \sigma^{\nu}_{\rm HI} \int_0^t dt' \dot{N}^{\nu}_{\rm ion}(t', \nu') \left[ c\kappa_\nu e^{-\int_{t'}^{t} dt'' c\kappa_{\nu''}} \right] \tag{23}$$

The term in brackets is $G(t, t', \nu, \nu')$ (Equation (5)). It is straightforward to see that Equation (23) is the result of directly substituting Equation (5) into Equation (4), and then putting the result into Equation (3).

## Appendix 2. Insensitivity of $\dot{N}_{\rm ion}$ measurements to neutral islands

In Figure 7, we show the ratios of the black dotted, red dashed, and cyan dot-dashed curves in Figure 2 with the maximum likelihood fiducial measurement (black solid curve). We see that during reionisation, accounting for neutral islands (cyan curve) changes the measurement by at most 15% while reionisation is ongoing, and only at the couple-percent level afterwards. This is smaller than the error incurred by neglecting redshifting ($\approx$ 20%) or using the LSA (a factor of 2) at $z = 2.5$. This is despite the fact that our assumed reionisation scenario is somewhat extreme, such that we likely over-estimate the importance of islands for the measurement at $5 < z < 6$.

We can understand the reason for this result by considering the physical meaning of the MFP. In Appendix C of Cain, D'Aloisio, et al. (2024), we showed that the frequency-averaged MFP is well-approximated by

$$\lambda^{-1} = \frac{\langle \Gamma_{\rm HI} n_{\rm HI} \rangle_V}{F_\gamma} \tag{24}$$

where the numerator is the volume-averaged absorption rate in the IGM and $F_\gamma$ is the average incident ionising photon flux. The former is simply $\dot{N}_{\rm abs}$, and the latter is given by $F_\gamma = N_\gamma c$, where $N_\gamma$ is the mean number density of ionising photons. So, we can write

$$\lambda^{-1} = \frac{\dot{N}_{\rm abs}}{N_\gamma c} \tag{25}$$

That is, the IGM absorption coefficient is the absorption rate divided by the ionising flux. It is straightforward to define "ionised" and "neutral" components of Equation 25 such that

$$\lambda^{-1} = \lambda^{-1}_{\rm ionised} + \lambda^{-1}_{\rm neutral} = \frac{\dot{N}_{\rm abs,ionised}}{N_\gamma c} + \frac{\dot{N}_{\rm abs,neutral}}{N_\gamma c} \tag{26}$$

Indeed, the first equality is just Equation 7 and the second is Equation 8. It follows that treating the IGM absorption as one component (Equation 24) or two components (ionised and neutral, Equation 25) should give the same result, provided $\lambda$ is appropriately defined. Of course, it is unclear whether the quantity recovered by the observations precisely matches the "theoretical" definition of the MFP (Roth et al. 2024). Still, this shows the physical reason why we should not expect treating absorption by neutral islands separately to significantly affect our measurement.

In the top panel of Figure 8, we break down the contribution of the different components of the IGM (ionised and neutral) to the total absorption rate, $\dot{N}_{\rm abs}$ (Eqs. (4) and (8)), in our fiducial measurement. The black solid curve shows $\dot{N}_{\rm abs}$ without accounting for neutral islands (Equation (4)), and the cyan dashed curve shows the same accounting for islands (Equation (8)). The red dotted and green dot-dashed curves show contributions to $\dot{N}_{\rm abs}$ in this case arising from only ionised gas ($\dot{N}_{\rm abs,ionised}$), and from only neutral islands ($\dot{N}_{\rm abs,neutral}$), respectively. Although the measured absorption by ionised gas when including the effects of islands is smaller than it is without including them, the additional absorption by the islands almost offsets this difference, yielding nearly the same total $\dot{N}_{\rm abs}$. Another interesting result is that $\dot{N}_{\rm abs,neutral}$ is always less than half of $\dot{N}_{\rm abs,ionised}$, even at $z = 6$ when the IGM is 30% neutral in our assumed reionisation model. Our fiducial measurement therefore suggests that the majority of absorption during reionisation's tail end takes place in the highly ionised IGM – the so-called "absorption-dominated" scenario (Davies et al. 2021; Davies, Bosman, and Furlanetto 2024).

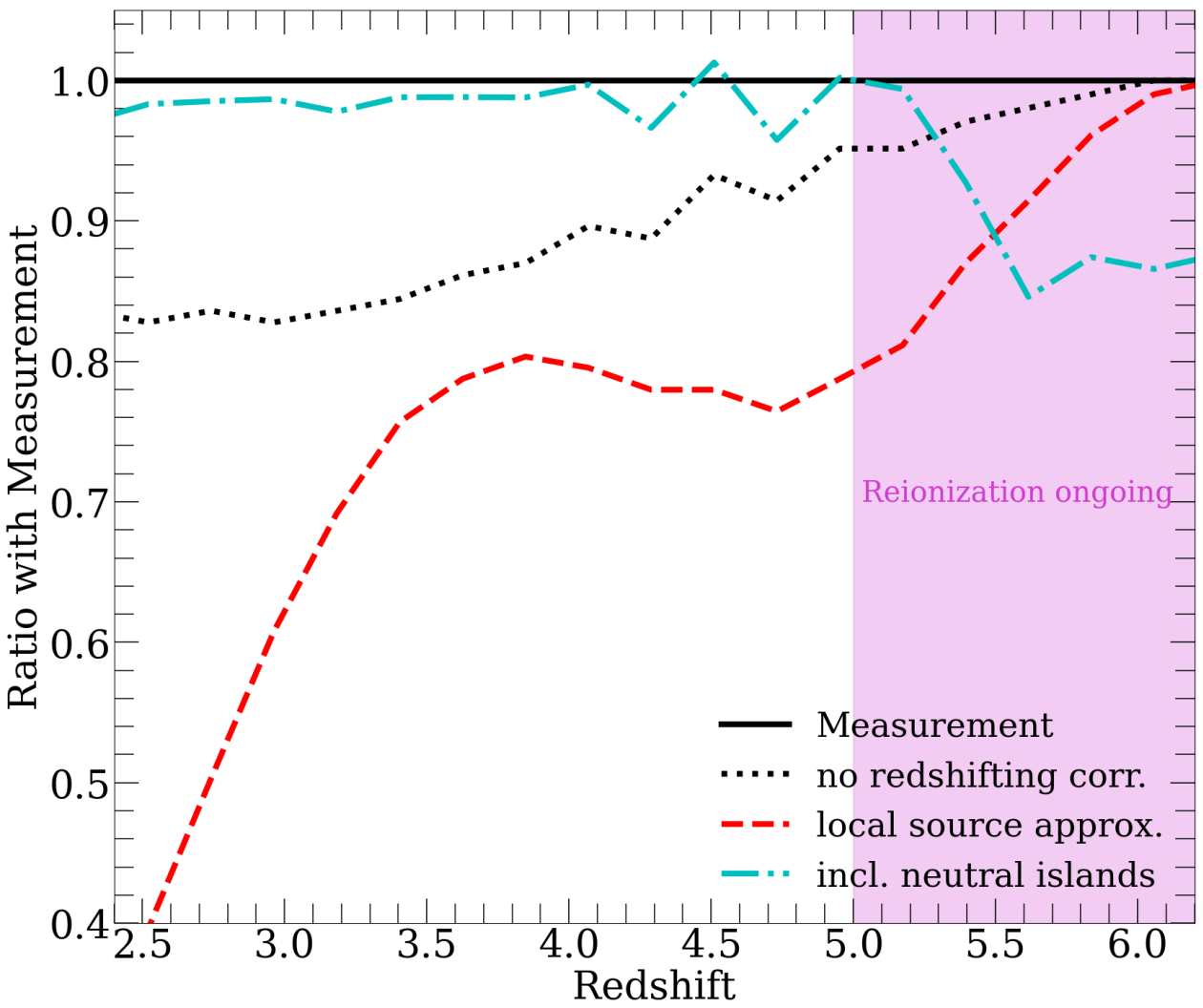

**Figure 7.** Relative effect on our measurement of neglecting redshifting (black dotted curve), using the LSA (red dashed curve) and accounting for neutral islands at $z > 5$ (cyan dot-dashed curve). The effect of accounting for islands is smaller during reionisation than that of the other two effects at $z \approx 2.5$, even for the somewhat extreme reionisation scenario assumed here.

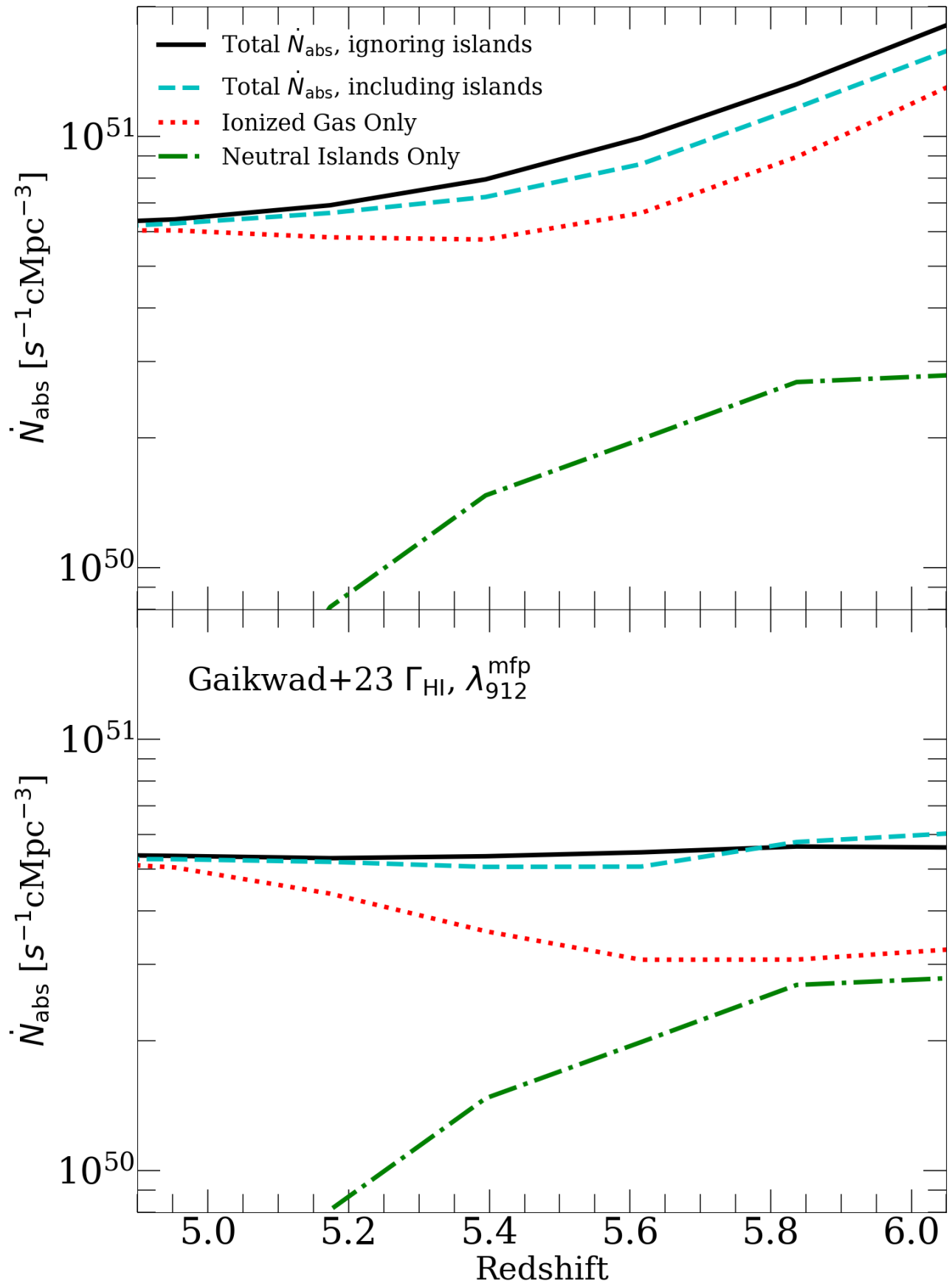

**Figure 8.** Contribution to the total absorption rate by the ionised IGM and neutral islands. **Top**: The black solid curve shows $\dot{N}_{\rm abs}$ for our fiducial measurement without accounting for neutral islands at $z > 5$, and the cyan dashed curve shows the same accounting for islands. The red dotted and green dot-dashed curves show the absorption rate by only ionised, and only neutral gas, respectively. **Bottom**: the same, but for our G23-only measurement. In this case, $\dot{N}_{\rm ion,ionised}$ and $\dot{N}_{\rm ion,neutral}$ are comparable at $z = 6$, but still add to a value close to the black solid curve. This shows that the standard formalism works well even if absorption by ionised gas does not dominate the total budget.

In the bottom panel, we show the same thing, but for our G23-only measurements. In this case, $\dot{N}_{\rm abs,ionised} \approx \dot{N}_{\rm ion,neutral}$ at $z \geq 5.8$, such that reionisation is not absorption-dominated. However, even in this case, the total is very close to the black solid curve. This suggests that even if the absorption rate is not dominated by ionised gas, the standard formalism remains a good approximation. This is good news for measurements of $\dot{N}_{\rm ion}$ using this technique, since the reionisation history at $5 < z < 6$ is poorly constrained observationally.

## Appendix 3. Fitting functions for $\Gamma_{\rm HI}$ and $\lambda_{912}^{\rm mfp}$

Here, we give the maximum likelihood fits to the sets of $\Gamma_{\rm HI}$ and $\lambda_{912}^{\rm mfp}$ measurements used in our analysis. Our fiducial result (including all measurements) is shown in Figure 1. We show fits to the two subsets described in the main text in Figure 9, in the same format as Figure 1. The left column shows our fit using only G23 data at $z > 5$, and the left panel excludes the G23 at these redshifts. In the first case, $\Gamma_{\rm HI}$ decreases with redshift more rapidly than in the fiducial fit, and $\lambda_{912}^{\rm mfp}$ declines less rapidly. The opposite is true in the right column. In this case, the $z > 5$ evolution of $\lambda_{912}^{\rm mfp}$ is determined entirely by the direct measurements from Zhu et al. (2023).

We fit the $\Gamma_{\rm HI}$ measurements, in units of $\rm s^{-1}$, to a fifth-order polynomial of the form

$$\log_{10}(\Gamma_{\rm HI}/[\rm s^{-1}]) = \sum_{n=0}^{4} a_n z^n \tag{27}$$

where $(a_0, a_1, a_2, a_3, a_4) = (-0.47, -11.50, 4.09, -0.62, 0.03)$ are the maximum likelihood parameters for our fiducial fit. Using only G23 data at $z > 5$, we obtain $(a_0, a_1, a_2, a_3, a_4) = (-6.52, -5.08, 1.62, -0.20, 0.01)$, and excluding the G23 measurements, $(a_0, a_1, a_2, a_3, a_4) = (-5.18, -6.60, 2.24, -0.32, 0.02)$.

Our fit to $\lambda_{912}^{\rm mfp}$ measurements, in units of $h^{-1}$cMpc, has the functional form

$$\log_{10}\left([\lambda_{912}^{\rm mfp} + 1]/[h^{-1}{\rm cMpc}]\right) = \frac{a_1 z^{b_1}}{1 + (z/a_2)^{b_2}} \tag{28}$$

where $(a_1, b_1, a_2, b_2) = (3.95, -0.46, 5.92, 13.90)$ is the maximum likelihood result for our fiducial fit. Using only G23 measurements at $z > 5$, we find $(a_1, b_1, a_2, b_2) = (3.98, -0.46, 6.08, 13.38)$, and without those measurements, we find $(a_1, b_1, a_2, b_2) = (3.98, -0.46, 5.77, 16.77)$. We emphasise that these fitting functions should not be extrapolated much beyond $2.5 < z < 6$.

## Appendix 4. Error budget for $\dot{N}_{\rm ion}$

Here, we take a closer look at the breakdown of our error budget, and compare our errors more closely to those of previous works that measure $\dot{N}_{\rm ion}$. Figure 10 shows our fiducial measurement with errors from the top panel of Figure 2, and compares this to the errors in Becker and Bolton (2013), which are several times larger than ours. In that work, the authors combined statistical and systematic errors from a number of sources to measure $\dot{N}_{\rm ion}$, including uncertainty in $\alpha$ and $\beta_{\rm N}$ as we do in this work (see their Table 3). They combined uncertainties from difference sources by linearly adding the logarithmic errors from each component, which assumes maximal correlation between sources of error and is the most conservative assumption. Our analysis assumes independence between different sources of error. If we make the same assumption using the reported errors in Becker and Bolton (2013), and add their fractional errors in quadrature, we recover the magenta points, which have $\pm 1\sigma$ ranges about half as large as their reported errors.

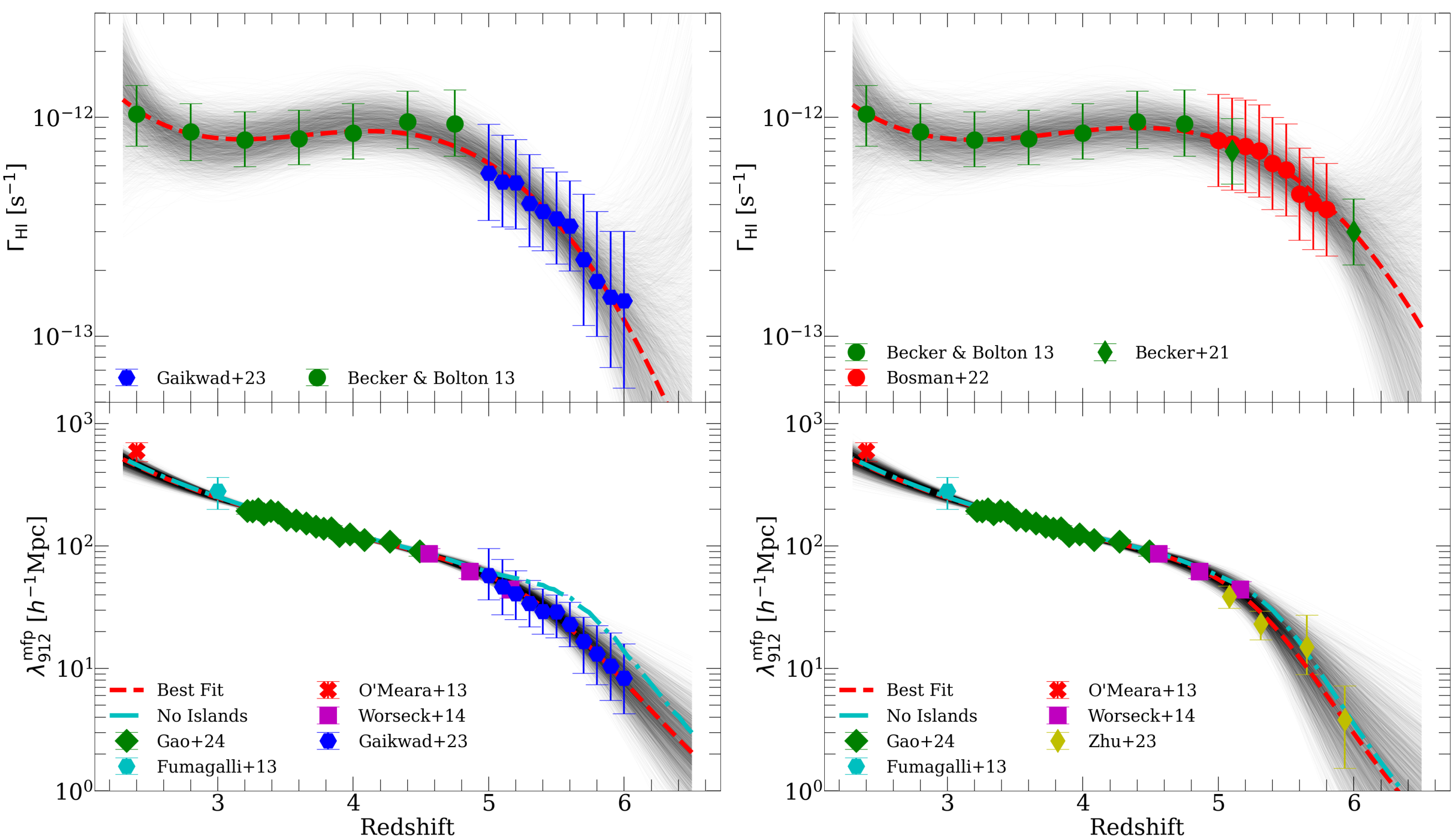

**Figure 9.** Fits to alternative sets of $\Gamma_{\rm HI}$ and $\lambda_{912}^{\rm mfp}$ measurements. **Left column:** fit using only G23 measurements at $z > 5$, in the same format as Figure 1. **Right column:** the same, but excluding only the G23 points at $z > 5$. In the left column, we see faster (slower) redshift evolution in $\Gamma_{\rm HI}$ ($\lambda_{912}^{\rm mfp}$) than in the fiducial case, and the opposite is true in the right column. This has a significant effect on inferred galaxy ionising properties at $z > 5$, as shown in the main text.

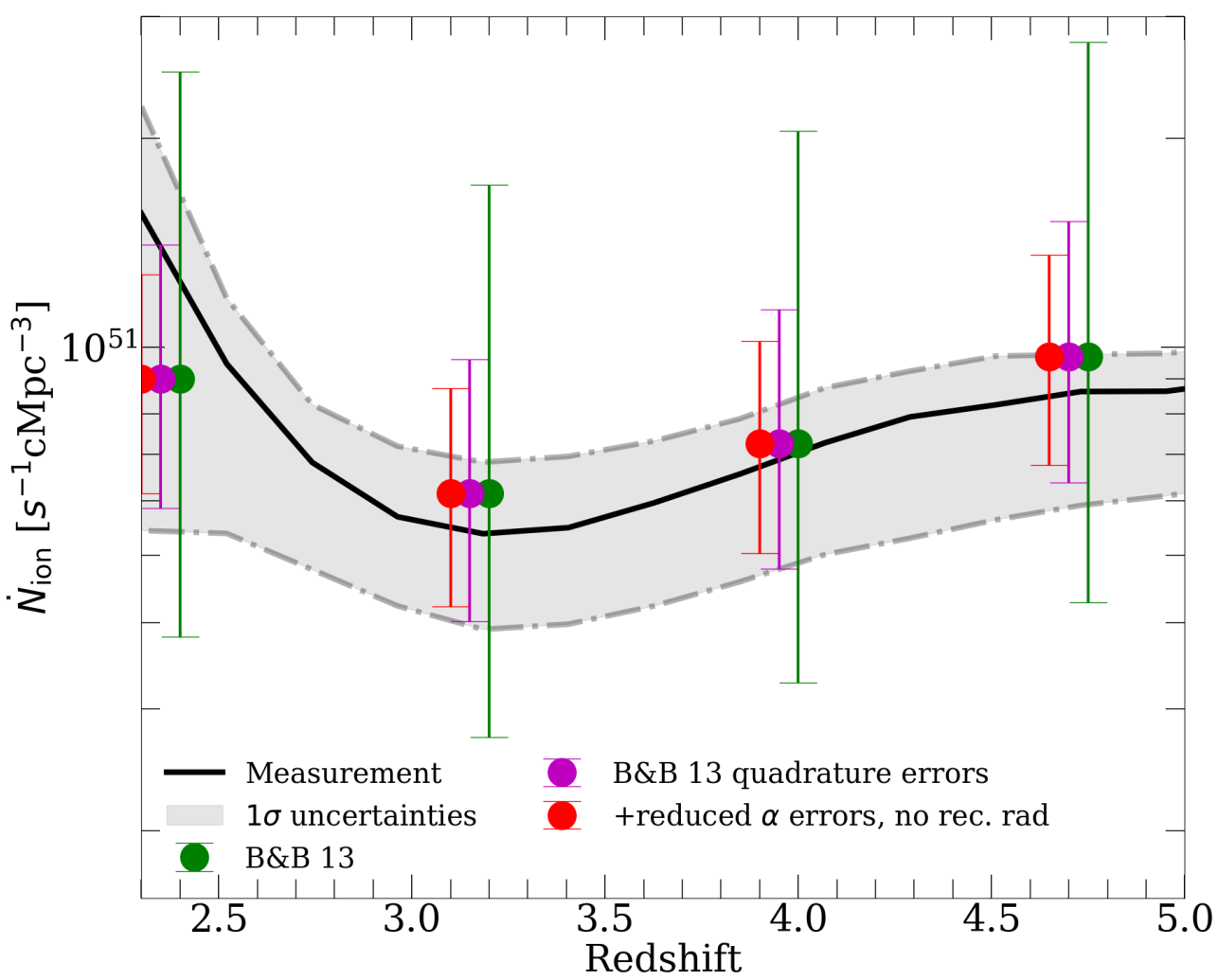

**Figure 10.** A more careful comparison of our error bars to those of Becker and Bolton (2013) at $z < 5$. We show our fiducial measurement and error bars from the top panel of Figure 2, alongside the reported Becker and Bolton (2013) measurements and error bars (green points) and two revisions of their errors. The purple points show what happens if the error from different sources in their analysis are combined in quadrature rather than linearly, which is more consistent with the assumptions made in our work. The red points further adjust the size of some of their systematic errors to better reflect those in our work (see text). With these revisions, their errors are only $\approx 30\%$ larger than ours - the rest of the difference is likely explained by lack of flexibility in our parametric fits to $\Gamma_{\rm HI}$ and $\lambda_{912}^{\rm mfp}$ measurements.

There are a couple additional differences between our analysis and that of Becker and Bolton (2013) can explain some of the remaining difference between our error bars and theirs. First, they do not assume a 4 Ryd cutoff in their ionising spectrum, which results in their uncertainties from $\alpha$ being larger than ours. They also include a systematic uncertainty for the effects of recombination radiation that is not included in our analysis. If we further adjust their errors to reflect these differences, we get the red points, which have uncertainties only $\approx 30\%$ larger than ours. The remaining difference is likely explained by the fact that we fit a parametric function to a large number of $\Gamma_{\rm HI}$ measurements, which may be artificially constricting our uncertainties. Another factor is that we do not include the full error covariance matrix for $\Gamma_{\rm HI}$ measurements from Becker and Bolton (2013) (which we note is unavailable in Bosman et al. (2022) and Gaikwad et al. (2023)). Despite these factors, the relatively good agreement between the red

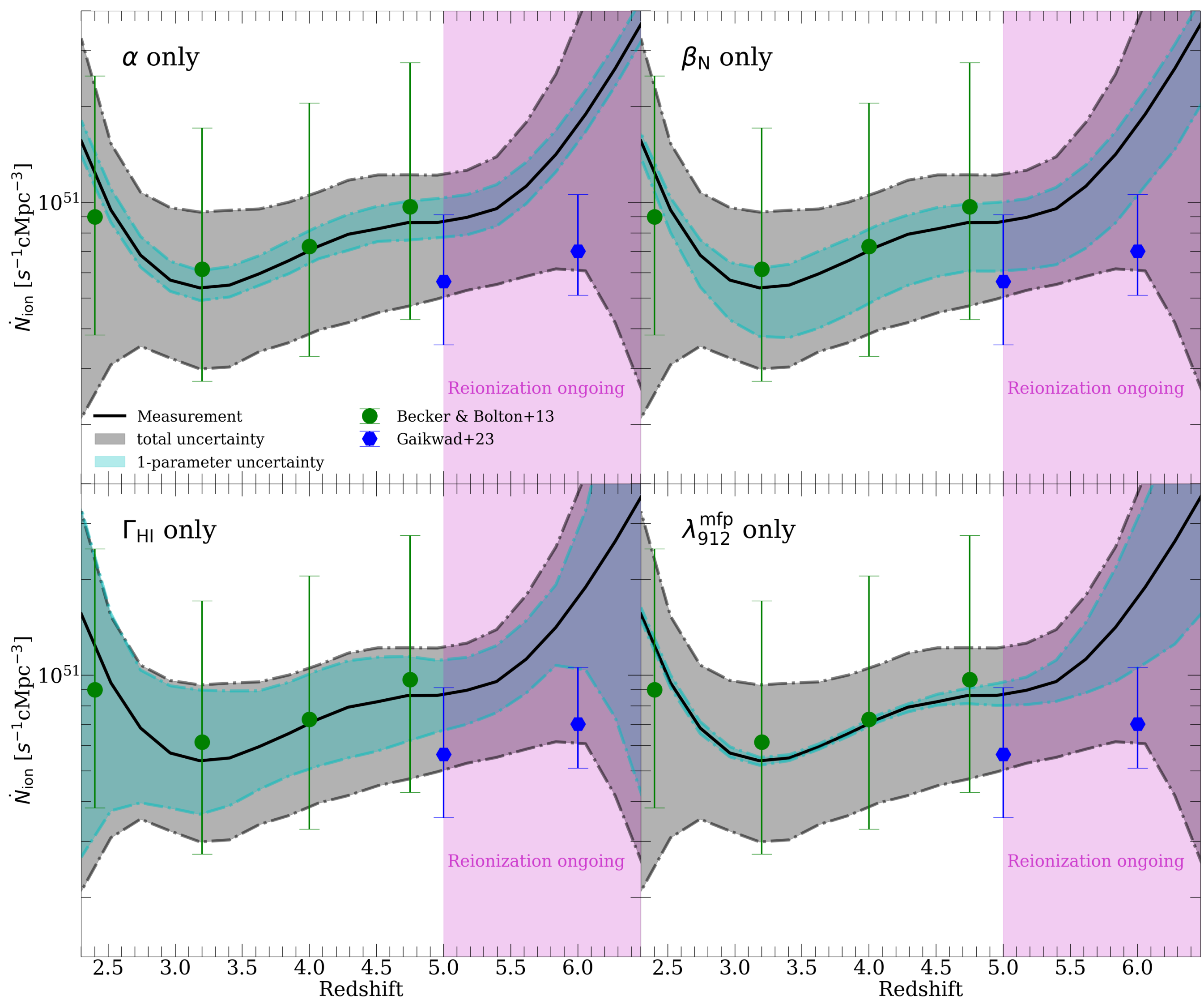

**Figure 11.** Breakdown of the contribution to the total error budget for $\dot{N}_{\rm ion}$ from each source of uncertainty. Each panel shows the full errors (black curve and shaded region, $2\sigma$) and the contribution from one parameter at a time. We see that $\Gamma_{\rm HI}$ and $\beta_{\rm N}$ are comparable and dominate sources of error at most redshifts, and $\alpha$ is always sub-dominant. Uncertainties in $\lambda_{912}^{\rm mfp}$ are only important at $z > 5$.

points in Figure 10 and our errors is encouraging.

In Figure 11, we show the full breakdown of contributions to our $1\sigma$ errors across all redshifts from uncertainty in $\alpha$, $\beta_{\rm N}$, $\Gamma_{\rm HI}$, and $\lambda_{912}^{\rm mfp}$ (see Table 1). In each panel, we show the fiducial fit and total uncertainty ($2\sigma$) compared to the error arising from varying only one parameter at a time. As seen from Table 1, $\beta_{\rm N}$ and $\Gamma_{\rm HI}$ dominate the uncertainty budget at most redshifts, and $\alpha$ is always sub-dominant. Uncertainties from $\lambda_{912}^{\rm mfp}$ are negligible at $z < 5$, but become important at $z > 5$ and are comparable to the other major sources of uncertainty at $z = 6$.